\documentclass[sigconf]{acmart}

\copyrightyear{2025}
\acmYear{2025}
\setcopyright{cc}
\acmConference[ASSETS '25]{The 27th International ACM SIGACCESS Conference on Computers and Accessibility}{October 26--29, 2025}{Denver, CO, USA}
\acmBooktitle{The 27th International ACM SIGACCESS Conference on Computers and Accessibility (ASSETS '25), October 26--29, 2025, Denver, CO, USA}
\acmDOI{10.1145/3663547.3746372}
\acmISBN{979-8-4007-0676-9/2025/10}

\usepackage{array}
\usepackage{subcaption}
\usepackage{url}
\newcolumntype{R}[1]{>{\raggedright\arraybackslash}p{#1}}

\begin{document}

\title[VeasyGuide]{VeasyGuide: Personalized Visual Guidance for Low-vision Learners on Instructor Actions in Presentation Videos}


\author{Yotam Sechayk}
\affiliation{%
  \institution{The University of Tokyo}
  \city{Tokyo}
  \country{Japan}
}
\orcid{0009-0002-5286-0080}
\email{sechayk-yotam@g.ecc.u-tokyo.ac.jp}

\author{Ariel Shamir}
\affiliation{%
  \institution{Reichman University}
  \city{Herzliya}
  \country{Israel}}
\orcid{0000-0001-7082-7845}
\email{arik@runi.ac.il}

\author{Amy Pavel}
\affiliation{%
  \institution{University of California, Berkeley}
  \city{Berkeley}
  \state{CA}
  \country{USA}}
\orcid{0000-0002-3908-4366}
\email{amypavel@berkeley.edu}

\author{Takeo Igarashi}
\affiliation{%
  \institution{The University of Tokyo}
  \city{Tokyo}
  \country{Japan}
}
\orcid{0000-0002-5495-6441}
\email{takeo@acm.org}

\renewcommand{\shortauthors}{Sechayk et al.}

\begin{abstract}
    Instructors often rely on visual actions such as pointing, marking, and sketching to convey information in educational presentation videos. These subtle visual cues often lack verbal descriptions, forcing low-vision (LV) learners to search for visual indicators or rely solely on audio, which can lead to missed information and increased cognitive load. To address this challenge, we conducted a co-design study with three LV participants and developed VeasyGuide, a tool that uses motion detection to identify instructor actions and dynamically highlight and magnify them. VeasyGuide produces familiar visual highlights that convey spatial context and adapt to diverse learners and content through extensive personalization and real-time visual feedback. VeasyGuide reduces visual search effort by clarifying what to look for and where to look. In an evaluation with 8 LV participants, learners demonstrated a significant improvement in detecting instructor actions, with faster response times and significantly reduced cognitive load. A separate evaluation with 8 sighted participants showed that VeasyGuide also enhanced engagement and attentiveness, suggesting its potential as a universally beneficial tool.
\end{abstract}




\begin{teaserfigure}
    \includegraphics[width=\textwidth]{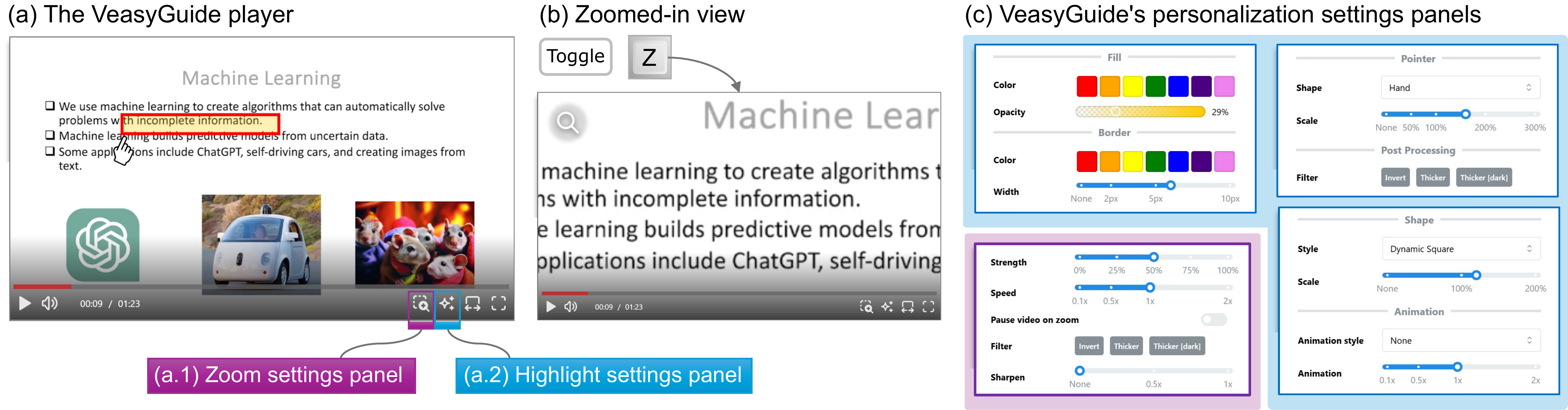}
    \caption{VeasyGuide’s interface includes: (a) a video player with highlighted areas (red rectangle and hand pointer), (a.1) zoom settings toggle, (a.2) highlight settings toggle, (b) a zoomed window (toggled with the Z key), and (c) zoom and highlight settings panels. While watching a video, VeasyGuide auto-highlights pointing, marking, and sketching activities. Users can zoom into highlighted portions with the Z key, adjust zoom with arrow keys, and customize highlight and zoom appearance in real time.}
    \Description{Illustration of the VeasyGuide player and its settings. From left to right. First, the player interface showing a video of a "Machine Learning" slide, where the phrase "incomplete information" is highlighted with a red rectangle, and standard video controls are visible below; with two additional buttons "Highlight personalization panel," and "Zoom personalization panel." Then, illustration of the z-key with the text "Toggle", atop a magnified view of the highlighted portion of the slide with the phrase "incomplete information" at the center. Finally, multiple personalization option panels.}
    \label{fig:interface}
\end{teaserfigure}

\maketitle

\section{Introduction}
\label{sec:introduction}

Presentation videos play a central role in modern education, spanning both formal and informal learning environments~\cite{navarrete2023videoLearning, wajdi2023blackboardApp, gosh2023tiktokLearning}. These presentations frequently incorporate slides enriched with visual aids such as videos~\cite{hartsell2006video}, figures~\cite{peng2021Slidecho}, and animations~\cite{grice2009can} to convey key concepts. Instructors often interact with slides in real-time using digital pens or pointers to direct viewer attention, annotate slide elements, or add new dynamic content~\cite{navarrete2023videoLearning}, typically in coordination with their speech.

However, low-vision (LV) learners, who experience limited visual access, often struggle to perceive these visual interactions as they occur, leading to missed details. Although universal design for learning and accessible presentation guidelines~\cite{burgstahler2010universal, ladner2017making, d2021accessible} recommend verbally describing all visual content, instructors frequently omit such descriptions. For example, instructors may use demonstrative pronouns (``this'') without clarifying the referent, or fail to mention when marking key elements with digital pens. In some cases, instructors do not verbalize complex spatial or relational information when adding sketches to slide content~\cite{d2021accessible}. For LV learners, identifying and interpreting these \textit{visual activities} constitutes a visual search problem, which increases cognitive load and hinders comprehension~\cite{zhao2016cuesee}.

Similar to prior work~\cite{peng2021Slidecho}, we analyzed 300 presentation videos across five domains to examine the frequency of visual activities. We define presentation videos as those that feature screen-shared slides, optionally accompanied by a video of the instructor. Visual activities were present in all domains, with 20\% to 50\% of videos containing at least one such activity (Appendix~\ref{apdx:presentations}). The highest proportions were in \textit{Natural Sciences} (50\% of videos) and \textit{Formal Sciences} (43.3\% of videos). We also examined the types of visual activities used. In \textit{Natural Sciences}, sketching and pointing were more frequent. In \textit{Formal Sciences}, sketching was more frequent, while in \textit{Social Sciences}, marking appeared more frequently than sketching or pointing. These findings show that access to visual activities---pointing, marking, and sketching---is essential for equitable access for LV learners.

To address this gap and support visual access for LV learners, we developed \textbf{VeasyGuide}, co-designed with 3 LV participants. First, we explored how participants perceive visual activities in educational presentation videos. Then, collaborated with participants to iteratively design VeasyGuide, addressing challenges associated with visual search. VeasyGuide enhances presentation videos by highlighting visual activities during playback, and enabling magnification that transitions automatically on new activities. The system has two components: (1) activity recognition and (2) activity visualization. The activity recognition pipeline employs computer vision motion detection techniques~\cite{parveen2021motion} and uses a graph-based representation to identify instructor interactions. Then, the activity visualization module augments video playback with personalized highlights and zooming capabilities.

To assess VeasyGuide’s effectiveness, we conducted a user study with 8 LV participants, evaluating their ability to detect visual activities and examining their interaction with presentation videos. We used real-world presentation videos featuring an instructor’s voice and screen-shared slides, where the instructor interacted directly with the slide content by pointing, marking, or sketching. Our findings show that VeasyGuide significantly increases the number of detected visual activities, with a mean of 88\% detection rate, compared with 61\%. VeasyGuide also improves the speed of detection, with a mean detection time of \( 1.53\) seconds, compared with \( 2.57 \) seconds. Additionally, using VeasyGuide results in reduced cognitive load and enhanced learning experience, which participants perceive as more accessible. To assess broader applicability, we also evaluated VeasyGuide on 8 sighted participants. We show that VeasyGuide improves engagement and attentiveness for sighted participants, highlighting its potential as a universally beneficial assistive feature, similar to how captions have become standard across media platforms. VeasyGuide is available online as a web application, visit \url{https://veasyguide.github.io/} for more information.

In summary, our contributions are:
\begin{itemize}
    \item A co-design study with LV participants to identify and address access barriers in presentation videos.
    \item VeasyGuide, a personalized tool that highlights and magnifies visual activities during playback.
    \item A user evaluation with LV participants, demonstrating the impact on accessibility of visual activities.
    \item A user evaluation with sighted participants, highlighting broad applicability.
\end{itemize}
\section{Background}
\label{sec:background}

VeasyGuide addresses visual search challenges posed by instructor interactions in educational presentation videos---a common barrier for low-vision (LV) learners.
We build on existing research in LV accessibility, content magnification, and personalization to improve access to visual activities in presentation videos.

\subsection{Video Accessibility for Visual Impairments}

Traditionally, video accessibility for blind and visually impaired users is achieved through sensory substitution~\cite{muhsin2024review}, most commonly via audio description (AD)~\cite{acb_audio_description_guidelines, snyder2005audio}.
AD provides descriptions of visual content~\cite{pavel2020rescribe, natalie2021efficacy}, and can be placed within the video time, or extend it~\cite{packer2015overview}.
Research in AD focused on improving the creation process~\cite{hirvonen2023co, pavel2020rescribe}, automating the creation altogether~\cite{wang2021toward, campos2023machine, han2023autoad, yuksel2020human, jiang2023beyond, Jain2023Sports}, and supporting AD consumption through spatial audio~\cite{Fan2023ImprovingAccessibility} or user exploration~\cite{ning2024spica, stangl2023potential}.

However, while LV users largely prefer to use their residual vision for visual content~\cite{szpiro2016people, jacko1998designing, wang2023understanding}, AD assumes no visual perception~\cite{acb_audio_description_guidelines}, therefore, not aligned with the needs of LV users.
Motivated by this, our approach focuses on enhancing visual access rather than substituting it.

\subsection{Accessibility for Low-vision Users}

For LV users, accessibility approaches emphasize sensory enhancement rather than substitution.
For instance, studies have explored using augmented reality (AR) to enhance real-world environments, demonstrating its potential to assist with daily tasks~\cite{zhao2020Designing}, support visual search~\cite{zhao2016cuesee}, aid navigation~\cite{ZhaoKRFA20}, and improve safety during activities such as stair use~\cite{ZhaoKCFA19} and obstacle avoidance~\cite{fox2023using}.
Others explore tangible enhancements through special equipment to support visually demanding activities like close-contact sports~\cite{TsutsuiYZSTO24}.

A common sensory enhancement strategy is content magnification.
Despite its long history, magnification software remains limited in functionality and design~\cite{tang2023screen}.
Recent work has proposed adaptive magnification techniques to support reading~\cite{WangPHKZM024}, browsing the web~\cite{billah18SteeringWheel}, operating mobile interfaces~\cite{islam23SpaceX}, and interpreting complex visuals such as diagrams~\cite{wang2024low, prakash2024understanding}.
Magnification has also been integrated into AR systems~\cite{stearns2018design, zhao2020Designing}.

For video content, prior work introduced post-processing filters to enhance visual clarity~\cite{sackl2020ensuring} and saliency-based magnification to reduce manual control~\cite{aydin2020videoMagnif}.
We extend this body of work by focusing on educational presentation videos.
Our system employs a two-part strategy: automatic detection of visually salient instructor interactions, and guided magnification using visual highlights as anchors.
This approach minimizes manual effort and enhances the viewing experience for LV users.

\subsection{Accessibility of Visual Content in Presentation Videos}

Video presentation are widely used in both formal and informal education~\cite{navarrete2023videoLearning}, appearing across platforms such as YouTube~\cite{youtube}, Khan Academy~\cite{khanacademy}, Coursera~\cite{coursera}, and live conferencing software like Zoom~\cite{zoom}.
These videos often feature slides, screen-shared tutorials, and interactive blackboards.

Prior work has explored tools to help presenters create more accessible content~\cite{peng2021say}, author accessible slides~\cite{elias2018slidewiki}, and produce immersive tutorials with accessibility in mind~\cite{kong2021tutoriallens}.
Despite these efforts, presentation videos remain challenging for LV learners~\cite{Jiang2024Context, schaadhardt2021understanding}.
Visual content may be too small, beyond the current field of view, or insufficiently described~\cite{sechayk2024smartLearn, jung2018dynamicslide}.
Additionally, the cognitive load of simultaneously processing verbal and visual information can hinder learning~\cite{hinojosa2015investigations, szpiro2016people}.

To address these issues, researchers have explored techniques such as highlighting text segments aligned with speech~\cite{jung2018dynamicslide, yip2021visionary}, identifying slide elements not referenced by the speaker~\cite{peng2021Slidecho}, and enabling user-guided magnification based on segmented slide elements~\cite{denoue2013real}.
While instructors frequently use pointing, sketching, or marking to support learning~\cite{d2021accessible}, these interactions are often overlooked in accessibility solutions.
Our work addresses this gap by enhancing the visibility of instructor interactions through visual highlights and integrated magnification.

\subsection{Personalization and Context-Aware Accessibility}

As accessibility research evolves, the limitations of \textit{one-size-fits-all} solutions have become increasingly apparent~\cite{stangl2021going, Jiang2024Context, stangl2023potential, chang2024worldscribe}.
Personalization is not only beneficial but, in many cases, essential, particularly for LV users, as they can have vastly different levels of visual perception, leading to significantly different visual experiences.
For example, \citet{wang2024low} highlight diverse screen magnifier usage across users, while \citet{zhao2020Designing} emphasize the need for personalized interaction design.

Furthermore, accessibility needs and preferences often vary by context.
For instance, \citet{stangl2021going} show that image description needs may differ across scenarios, and \citet{chang2024worldscribe} proposes a context-aware description system for real-world scenes.
Similarly, \citet{natalie2024audio} argue for customizable AD, and \citet{Jiang2024Context} demonstrate that AD preferences shift across video genres.
Even preferences for post-processing filters that enhance visual clarity are highly individual~\cite{sackl2020ensuring, billah18SteeringWheel}.

To support personalization, researchers have developed user-guided tools~\cite{chang2024worldscribe, stangl2023potential, peng2021Slidecho, natalie2024audio}, mapped design spaces through user studies~\cite{wang2024low, stangl2021going, Jiang2024Context}, and adopted participatory design approaches~\cite{aydin2020videoMagnif, TsutsuiYZSTO24, jiang2022co, mattheiss2017user}.
Our work aligns with this trajectory.
Through a participatory co-design study, we incorporated feedback from LV users to inform VeasyGuide’s development.
We prioritize personalization, which is critical for LV users due to their varying degrees of residual vision and context-dependent needs.
VeasyGuide customizes visual highlights and in-video magnification of instructor interactions, which, to the best of our knowledge, has not been previously explored in this context.
\section{Co-design Study}
\label{sec:co_design}

We conducted a co-design study with 3 low-vision (LV) participants to investigate the following design questions in the context of educational presentation videos:
\begin{itemize}
    \item[\textbf{DQ1:}] How do LV learners experience visual activities?
    \item[\textbf{DQ2:}] What challenges do visual activities introduce?
    \item[\textbf{DQ3:}] What makes visual activities easier to detect and follow?
\end{itemize}

In the first session, we explore participants' experiences with visual activities in presentation videos (DQ1) and the challenges they present (DQ2). Then, in sessions 2--4, we iteratively refine VeasyGuide (DQ3). Finally, we dedicate session 5 to reflect on the co-design process itself from the perspective of the participants.
Since the first author is an individual with low vision, we acknowledge their influence throughout the co-design study. For instance, the initial prototype was based on both the experience and challenges of the participants, and the lived experiences of the first author.

\subsection{Method}

We recruited three LV participants, one male (aged 36) and two female (aged 30 and 33) through personal networks.
Participants had varying visual conditions that resulted in diverse accessibility needs (see Table~\ref{tab:codesign_participants}).
The study was conducted remotely via the Zoom~\cite{zoom} video conferencing platform, consisting of five 1-hour sessions per participant, held individually over five weeks.
Participants were compensated with a \$100 USD gift card for their full participation.
We used a mixture of educational presentation videos provided by the participants, along with a video that we supplied. This approach ensured consistency in the content viewed by participants while exploring real-world scenarios they encountered prior. The video we provided was created by a professional instructor who was compensated at their hourly rate.

\begin{table*}
    \centering
    \begin{tabular}{ccclllc}
    \toprule
    \textbf{PID} & \textbf{Age} & \textbf{Gender} & \textbf{Diagnosed Condition}    & \textbf{Onset} & \textbf{Education Level} & \textbf{Screen Magnifier} \\ \midrule
    C1               & 36           & Male            & Albinism, astigmatism, photophobia, nystagmus & Congenital     & HS              & Yes                 \\
    C2                & 30           & Female          & Retinitis pigmentosa, narrow FoV, limited night vision  & Congenital     & UG            & No                     \\
    C3              & 33           & Female          & Duane syndrome        & Congenital     & UG            & No                     \\
    \bottomrule
\end{tabular}
    \caption{Details on co-design participants. Education level abbreviations: UG (Undergraduate), HS (High School). \emph{FoV} abbreviates field of vision.}
    \label{tab:codesign_participants}
\end{table*}

\subsection{Findings}

\subsubsection{Experience and Challenges (DQ1--DQ2)}
All participants reported using presentation videos as a learning resource and described missing visual information as an unavoidable aspect of their experience.
C1 remarked, \textit{``Not knowing that I even missed something is a frustrating thing.''}
Since the appearance of visual activities can be inconsistent and not always intuitive, participants struggled with identifying \emph{what} to look for and \emph{where} to look, resulting in increased cognitive load and fatigue.
Consequently, they expressed that they often disregard visuals, leading to reduced engagement and missing important information.

Visual style significantly affected participants' ability to perceive visual activities.
Overall, reported challenges with visual styles included use of small pointers (C1), colors with low-contrast (C2, C3), and thin lines when using digital pens (all participants).
Specifically, C1 noted that marking actions were difficult to detect due to their subtlety and brevity, while C3 found that overlapping sketches blended together visually.
Moreover, we note how participants are divided in their preference of visual style, while C2 expressed she prefers dark content over bright backgrounds, C1 and C3 prefer the opposite. Therefore, visual enhancements should be configurable to various preferences.

To mitigate visibility challenges, some participants paused the playback or used techniques to magnify content. However, C2 and C1 noted that these strategies disrupts the flow of learning.
C1, a screen magnifier user, described the process of magnifying visual activities in videos as cumbersome, noting that it involves multiple manual operations.
Similarly, C2, who addresses the visual challenge by taking screenshots and enlarging them using image viewing software, explained that she often chooses to overlook unclear visuals.
C3 does not use content magnification and generally ignores unclear visuals.

\subsubsection{Accessibility Approach (DQ3)}
Insights from DQ1 and DQ2 indicated that, for LV learners, a distinct and personalized visual style for instructor actions is important. With magnification being integrated into the playback and not requiring tedious manual operation, aligning with prior work~\cite{aydin2020videoMagnif}. We began with a base prototype, which evolved into VeasyGuide through iterative refinement and user feedback.

\paragraph{Base Prototype}
We created the base prototype using the insights gained from DQ1 and DQ2, along with the lived experience of the first author, an individual with low vision.
To provide a distinct visual style, we chose an approach similar to \citet{sechayk2024smartLearn}, highlighting any notable visual change that happens on the video frame using a red semi-transparent circle with an adjustable diameter.
Since participants expressed the need for magnification, with existing options being cumbersome, we integrated an easy-to-operate zoom functionality into the video playback.
The zoom is anchored to entire visual activities, which are a sequence of consecutive visual changes, with a duration of $\geq 5$ seconds, and is toggled on or off when participants press the \verb|Z| key. We automatically zoom out when an activity is completed.

\paragraph{First Impressions}
All participants found the highlight useful---C1 remarked he \textit{``[feels] lost without it.''}
However, participants expressed concerns about the need for more control over visual features and the excessive motion caused by highlighting each visual change.
For instance, C1 and C3 found the red semi-transparent overlay challenging to see against darker backgrounds.
Furthermore, although zooming helped overall visibility, thin sketch lines remained difficult to discern.
C1 described the zooming as \textit{``rigid,''} and all participants desired more granular control and better integration between magnification and highlights.

\paragraph{Iterative Design Results}
We synthesized participants' feedback into three primary categories: \textbf{visual style}, \textbf{visual aids}, and \textbf{behavior}.
The final VeasyGuide interface is shown in Figure~\ref{fig:interface}, and key participant suggestions are summarized in Table~\ref{tab:iterative_feedback}.

\subparagraph{\textbf{Visual Style:}}
Participants preferred a bounding box over the original circular highlight.
While the circle offered a consistent shape, the box provided clearer spatial context.
C1 noted, \textit{``[the box] tells me exactly where and on what to look at.''}
To address participants’ desire for more control over the visual style of highlights, we incorporated personalization settings for fill color, opacity, and border styles.
Although we considered using adaptive styles based on activity type (pointing, marking, sketching), participants preferred a consistent visual appearance.

\subparagraph{\textbf{Visual Aids:}}
To make highlights easier to notice, we added pointer icons: either a cursor icon (\includegraphics[scale=0.04]{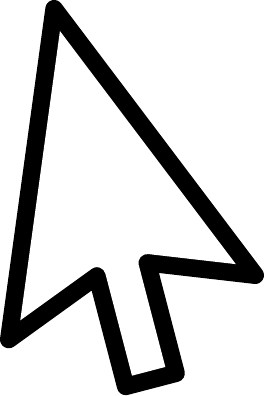}) or a hand icon (\includegraphics[scale=0.04]{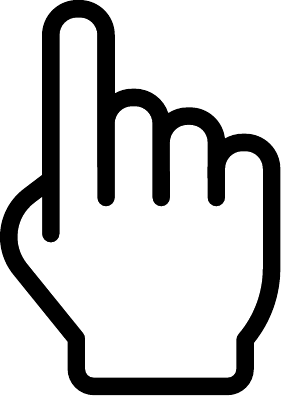}), both of which are familiar---easily recognizable---visual indicators.
C2 motivated these additions, \textit{``it is easier [to notice] when I know what to look for.''}
We also introduced optional visual filters to address issues with low contrast.
While exploring different visual filters is not the primary focus of this work, prior work demonstrates their usefulness for videos overall~\cite{sackl2020ensuring}. For VeasyGuide, we chose to include two filters---color inversion and line thickening---based on participant feedback.

\subparagraph{\textbf{Behavior:}}
To address the excessive motion of highlights, we transitioned to highlighting entire activities rather than each individual change. This approach results in a single stationary highlight defined by the activity’s start and duration.
However, C2 noted that some movement in the highlights helped guide her attention, so we added a size-variation animation.
C1 and C3 found this change uncomfortable or distracting, so we implemented the animation as an optional setting.
For instance, C1, who is photosensitive, experienced discomfort due to the fluctuations in brightness caused by the animation.
In place of animation, C1 and C3 suggested adding audio cues, aligning with prior work~\cite{peng2021Slidecho, Jiang2024Context, Jain2023Sports}. Audio cues are beyond the scope of this work and left for future exploration. However, we note their potential to increase cognitive load~\cite{Fan2023ImprovingAccessibility}.

To improve zoom integration with visual activities, as per participants' feedback, we changed how the zoom functions.
Instead of an abrupt zoom-out, the system now smoothly pans to the next activity, enhancing continuity.
Additional optional settings include the ability to automatically pause during zooming, as proposed by C2, and adjustable zoom animation speed, which C1 and C3 suggested.
Specifically, C2 expressed that \textit{``auto-pausing the video when zooming in gives me time to digest and adjust to the content, and the freedom to continue when I'm ready.''}
Finally, C1 and C2 emphasized the importance of anticipating instructor interactions to comfortably adjust their gaze. To address this, we added a 1.5-second pre-activity highlight trigger.

\begin{table}[t]
    \centering
    \begin{tabular}{R{2.4cm}R{3.6cm}l}
    \toprule
    \textbf{Suggestion}                            & \textbf{Motivation}             & \textbf{PIDs} \\
    \midrule
    \textbf{Consistent style}                     & Simpler visual search.          & C1, C2, C3      \\
    \hline
    \textbf{Post-processing filters}               & Clearer visuals.                & C1, C2, C3      \\
    \hline
    \textbf{Box shaped highlight}                   & Better spatial information.     & C1, C2, C3      \\
    \hline
    \textbf{Zoom control with auto pause}            & Content digestion.              & C1, C2             \\
    \hline
    \textbf{Advance notice of activities}          & Preparation time.               & C1, C2             \\
    \hline
    \textbf{Auto-transition of zoom}              & Reduced disruption.             & C1                  \\
    \hline
    \textbf{Animation}                             & Support narrow visual field. & C2                   \\
    \hline
    \textbf{Pointer}                               & Improved familiarity.        & C2                   \\
    \midrule
    \textbf{Audio cues} \textit{(not implemented)} & Notify to start visual search.        & C1, C3           \\
    \bottomrule
\end{tabular}
    \caption{Key suggestions from the iterative design process.}
    \label{tab:iterative_feedback}
\end{table}

\subsection{Co-design Reflections}
\label{sec:co-design_reflections}
All participants expressed satisfaction with their contributions and the representation of their needs in the final design.
C2 shared, \textit{``I was so happy to see animations [implemented] after I talked about it.''}
C1 remarked, \textit{``I can adjust what I really need, and adjust to different [video] contents.''}

Beyond system development, the co-design process had a broader impact on the participants.
Prior to the co-design study, C3 has avoided using accessibility tools due to her difficulties with adapting to different workflows.
She noted, \textit{``I tried using an iPad before because of the accessibility features, but it was very different from what I am used to, so I gave up on trying.''}
By the end of the study, C3 recognized the value of accessibility tools, and mentioned how her experience using and contributing to VeasyGuide motivated her to explore them further.

Moreover, although study sessions were conducted individually, features proposed by one participant often influenced others.
For example, after seeing the added pointer visuals suggested by C2, C1 recognized the importance of predictable visuals in supporting his ability to notice highlights, which sparked a broader discussion during the session.
The experience of C2 has indirectly resulting in C1 reflecting on his own visual perception.
This emphasizes the value of indirect peer influence in co-design, particularly for individuals with congenital sensory disabilities, who may lack comparative experiences and thus not question their current strategies.

Participants also noted challenges maintaining visual focus during sessions.
Sessions 2--4 required sustained screen engagement, leading to eye strain, discomfort and fatigue.
Although we invited participants to take breaks as needed, participants did not proactively suggest any breaks during our sessions.
When reflecting on their experience, participants suggested including predefined breaks in each session schedule to help reduce eye strain and discomfort.
They also proposed that break periods include explicit instructions to rest their eyes.
We recommend that future studies incorporate structured breaks to support LV participants’ visual endurance, along with clear guidance for eye rest.

\subsection{Final Design Implications}
\label{sec:design_implications}
Using insights from the co-design study, we outline four design implications (DIs) that informed the design of VeasyGuide to support visual search for LV users:

\begin{itemize}
    \item[\textbf{DI1:}] \textit{Visual Access.} Visual access should be enhanced through indicators that reduce the effort required for visual search.
    \item[\textbf{DI2:}] \textit{Familiarity.} By selecting a visual style, LV users become familiar with what they should expect, making visual cues easier to notice.
    \item[\textbf{DI3:}] \textit{Predictability.} Advance cues support user readiness and reduce stress.
    \item[\textbf{DI4:}] \textit{Personalization.} Tools must be configurable to accommodate the diverse needs, preferences, and contexts of LV users.
\end{itemize}

In VeasyGuide, we provide visual access to visual activities in educational presentation videos through highlights and magnification (\textbf{DI1}). The visual style of highlights, selected by users and consistent across different videos, creates a familiar visual cue that is easier to notice (\textbf{DI2}). By revealing highlights before the activity begins, along with spatial information conveyed using the box-shaped highlight, gaze direction and trajectory become more predictable (\textbf{DI3}). Finally, the suite of personalization features in VeasyGuide allows the experience to be configured to fit different users and to various video contents (\textbf{DI4}).

\section{VeasyGuide}
\label{sec:VeasyGuide}

We developed VeasyGuide through an iterative co-design process (Section~\ref{sec:co_design}).  The system comprises two main components: (1) an activity recognition pipeline (Section~\ref{sec:activity_recognition}), and (2) an activity visualization module (Section~\ref{sec:activity_visualization}) that augments video playback with personalized highlights and magnification.

\begin{figure*}[htbp]
    \centering
    \includegraphics[width=\textwidth]{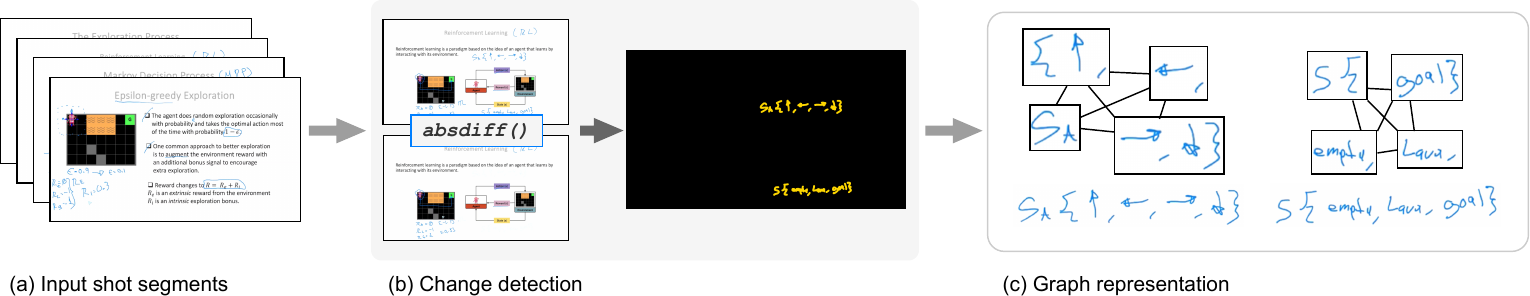}
    \caption{Overview of the VeasyGuide recognition pipeline. Input video segments are processed (a), and changes are detected using computer vision methods (b), then these changes are converted into a graph representation (c), where activities are detected and classified.}
    \Description{A flowchart showing the VeasyGuide recognition pipeline from left to right.}
    \label{fig:recognition}
\end{figure*}

\subsection{Activity Recognition Pipeline}
\label{sec:activity_recognition}

We implement the activity recognition pipeline in Python~\cite{python}, using OpenCV~\cite{opencv} for computer vision (CV) tasks and NetworkX~\cite{networkx} for graph-based algorithms. The pipeline consists of two stages: activity detection and activity classification (Figure~\ref{fig:recognition}). We employ CV motion detection techniques~\cite{parveen2021motion} to detect visual changes, which are then represented as a graph to identify and classify visual activities. Our approach is similar to that of \citet{ram2023vidadapter}.

\subsubsection{Activity Detection}
\label{sec:activity_detection}
We begin by splitting the input video into shots using an off-the-shelf shot detection library~\cite{pyscenedetect} to avoid slide transitions. Each shot is then divided into non-overlapping segments, with each segment containing up to one-third of a second’s worth of frames, based on the video’s frame rate. To detect visual changes, we compute the difference between the first and last frames in each segment. From the resulting frame difference, we extract contours~\cite{suzuki1985topological} using grayscale thresholding to highlight areas of change. We filter out small contours below a predefined area threshold (\(0.01\%\) of the frame area) to reduce noise caused by compression artifacts. The remaining contours are treated as Regions of Change (RoC).

Each RoC is represented as a node in a graph, defined by a tuple containing the timestamp of the first frame in the segment, the top-left corner coordinates of the bounding box, and the bounding box’s width and height. We connect nodes with an edge if their corresponding RoCs are both temporally and spatially close. RoCs are temporally close if their timestamps differ by less than a predefined threshold (\(3~\text{seconds}\)), and spatially close if the minimal distance between their bounding boxes is less than a predefined threshold (\(5\%\) of the frame diagonal). We then compute edge weights using Hu moments~\cite{hu1962visual} to define visual difference between RoC shapes.

To further reduce noise from transient object movements such as pointer trails~\cite{denoue2013real}, we merge RoC nodes that meet three criteria. First, their timestamps must be temporally adjacent, based on the set segment duration. Second, the regions must occupy the same position. Third, the visual difference between them must be less than a defined threshold (\(0.5\)), based on the Hu moments score.

\subsubsection{Activity Classification}
\label{sec:activity_classification}
We identify activities by analyzing the graph’s connected components (CCs), computing their bounding boxes and time spans for use in visual highlighting. Each activity is defined as a tuple that includes the CC of RoC nodes, the minimal bounding box that encloses all nodes in the component, and the earliest and latest timestamps among the nodes. While we considered a rule-based heuristic to classify activity types (e.g., pointing, marking, sketching), feedback from our co-design study (Section~\ref{sec:co_design}) indicated a preference for consistent visual styles. As a result, the current implementation omits activity-type classification, though we consider it a potential direction for future work.

\subsection{Activity Visualization}
\label{sec:activity_visualization}

VeasyGuide overlays visual highlights and magnification on top of instructional videos in real-time to support LV users in perceiving instructor activities such as pointing, sketching, or marking. We implemented the system as a web-based application using React~\cite{react} for the frontend and Flask~\cite{flask} for the backend. Users can interactively configure both visualization and magnification features to align with their individual visual needs.

\paragraph{Activity Highlighting.}
VeasyGuide automatically highlights detected instructor activities on the video. When multiple activities occur simultaneously, the system first filters for those that fully contain the current timestamp. If multiple matches remain, it selects the activity whose start time is nearest to the current time.

Users can customize highlight appearance and behavior through a settings menu, accessible via a button on the video player (\includegraphics[scale=0.4]{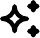}); see Figure~\ref{fig:interface}. The menu organizes settings into three categories: appearance, behavior, and enhancement features. For example, users can adjust border color and thickness to improve visibility against complex backgrounds or apply a post-processing filter to enhance the salience of sketch strokes. All changes apply immediately in a WYSIWYG\footnote{What You See Is What You Get} manner, allowing users to configure highlights in real time as the video plays. Table~\ref{tab:highlight_settings} lists all configurable highlight options and their default values.

\begin{table}[h]
    \centering
    \begin{tabular}{lll}
    \toprule
    \textbf{Category}   & \textbf{Setting} & \textbf{Default} \\ \midrule

    \textbf{Appearance} & Fill Color       & Yellow           \\
                        & Fill Opacity     & 15\%             \\
                        & Border Color     & Red              \\
                        & Border Width     & 4px              \\
\midrule
    \textbf{Behavior}   & Shape Style      & Box              \\
                        & Scale            & 100\%            \\
                        & Animation Style  & None             \\
                        & Animation Speed  & 1.0x             \\
\midrule
    \textbf{Enhancement}    & Pointer Style    & Hand             \\
                        & Pointer Scale    & 100\%            \\
                        & Filter           & None             \\
    \bottomrule
\end{tabular}
    \caption{Highlight settings with their default values.}
    \label{tab:highlight_settings}
\end{table}

\paragraph{Activity Magnification.}
Users can toggle magnification during playback by pressing the \verb|Z| key. When activated, VeasyGuide zooms into the most recent instructor activity and centers it within the viewport. To reduce visual clutter, the system temporarily hides the highlight when magnified. As new activities are detected, VeasyGuide smoothly repositions the view to follow them.

Users can adjust the zoom settings through a panel opened via a button on the video player (\includegraphics[scale=0.4]{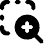}); see Figure~\ref{fig:interface}. The panel allows users to configure zoom strength, animation speed, and additional features such as auto-pause. For instance, enabling auto-pause causes the video to pause automatically whenever the view zooms into a new activity, giving users time to process visual information. Playback resumes either when users press play or zoom out. Table~\ref{tab:magnify_settings} lists all configurable magnification options and their default values.

\begin{table}[t]
    \centering
    \begin{tabular}{ll}
    \toprule
    \textbf{Setting} & \textbf{Default} \\ \midrule
    Strength    & 50\%             \\
    Speed       & 1.0x             \\
    Pause on Zoom    & False            \\
    Sharpness        & 1.0x             \\
    Filter           & None             \\
    \bottomrule
\end{tabular}
    \caption{Magnification settings with their default values.}
    \label{tab:magnify_settings}
\end{table}

\section{User Study}
\label{sec:user_study}

To evaluate VeasyGuide, we conducted a user study investigating its impact on accessibility for low-vision (LV) users and its universal applicability through exploration with sighted users.
Our user study addresses the following research questions (RQs) by comparing VeasyGuide with a baseline condition:
\begin{itemize}
    \item[\textbf{RQ1:}] Does VeasyGuide enable LV users to successfully locate a greater number of visual activities?
    \item[\textbf{RQ2:}] Does VeasyGuide reduce the time LV users spend searching for visual activities?
    \item[\textbf{RQ3:}] How does VeasyGuide alter the strategies LV users employ for visual search?
    \item[\textbf{RQ4:}] How does VeasyGuide impact the perceived accessibility and viewing experience of educational presentation videos for LV users?
    \item[\textbf{RQ5:}] What is the impact of using VeasyGuide on the viewing experience for sighted users?
\end{itemize}
The study includes two tasks: the first addresses RQ1--RQ3, and the second addresses RQ4--RQ5. Both sighted and LV participants completed all tasks using the same procedure. For both tasks, the baseline condition used an identical video player, with the highlight and magnification features disabled. The video player retained standard functionalities found in typical web-based players, such as the YouTube~\cite{youtube} player. After completing the tasks, participants took part in a semi-structured interview about their experience.

\subsection{Method}

\subsubsection{Participants}
We recruited 8 LV participants and 8 sighted participants (Table~\ref{tab:study_participants}).
LV participants were recruited through online message boards and organizations that support individuals with visual impairments. Eligible participants were 18 years or older, self-identified as having low vision, used their residual vision for reading, and had prior experience watching educational presentation videos.
Among the LV participants, three identified as male and five as female, with a mean age of 34.8 years (male mean: 33.3; female mean: 36.4).
Sighted participants were recruited through message boards at the authors’ institutions. Eligibility criteria included being 18 years or older and having prior experience with educational presentation videos.
Among the sighted participants, six identified as male and two as female, with a mean age of 27.8 years (male mean: 29.0; female mean: 24.5).

Each study participant received \$30 USD as compensation for their involvement in the 1.5-hour study session.
An additional \$10 USD was provided to one participant who encountered technical difficulties that extended their participation time.
Furthermore, recognizing professional expertise, one participant with a background in accessibility received an additional \$20 USD to align the compensation with their standard hourly rate.
This study has been approved by the Institutional Review Board (IRB) at the authors' institution.

\begin{table*}[htbp]
    \centering
    \begin{tabular}{lllR{7cm}R{1.5cm}R{1.5cm}R{1.5cm}c}
    \toprule
    \textbf{PID}                     & \textbf{Age} & \textbf{Gender} & \textbf{Diagnosed Condition}                                    & \textbf{Onset}           & \textbf{Screen Magnifier} & \textbf{Education Level} & \textbf{Monitor} \\
    \midrule
    L1                      & 31 & Male    & Left eye blindness, 10\% visual acuity, narrow FoV     & Congenital\ddag & Yes              & GR             & 32"     \\

    L2                      & 27 & Female  & Accommodative esotropia, Chiari malformation          & Congenital\ddag & Yes              & GR             & 24"     \\

    L3\textdagger           & 30 & Female  & Retinitis pigmentosa, narrow FoV, limited night vision & Congenital\ddag & No               & UG        & 27"     \\

    L4                      & 34 & Male    & Albinism, astigmatism, photophobia, nystagmus          & Congenital      & No               & GR             & 16"     \\

    L5                      & 41 & Female  & Albinism, astigmatism, photophobia, nystagmus                       & Congenital      & Yes              & UG        & 24"     \\

    L6 & 35 & Male    & Optic nerve hypoplasia                                 & Congenital      & Yes              & HS        & 32"     \\

    L7\textdagger           & 33 & Female  & Duane syndrome                                         & Congenital      & No               & UG        & 19"     \\

    L8                      & 51 & Female  & Diabetic retinopathy, glaucoma                         & Acquired        & Yes              & GR             & 27"     \\
    \midrule
    S1                      & 25 & Male    & Corrected vision                                       & Congenital      & Yes              & UG        & 14"     \\

    S2                      & 26 & Female  & Sighted                                                & Congenital      & No               & UG        & 16"     \\

    S3                      & 23 & Male    & Corrected vision                                       & Acquired        & No               & UG        & 14"     \\

    S4                      & 38 & Male    & Corrected vision, color blindness                      & Congenital      & Yes              & GR             & 15"     \\

    S5                      & 27 & Male    & Sighted                                                & Congenital      & No               & GR             & 29"     \\

    S6                      & 37 & Male    & Corrected vision                                       & Acquired        & No               & GR             & 14"     \\

    S7                      & 24 & Male    & Corrected vision                                       & Acquired        & No               & HS        & 13"     \\

    S8                      & 23 & Female  & Corrected vision                                       & Acquired        & No               & HS        & 13"     \\
    \bottomrule
\end{tabular}
    \caption{Information of study participants (L1--L8: LV participants, S1--S8: sighted participants). \emph{Monitor} indicates the size of the monitor used by participants during the study. Two participants marked with ~\textdagger~ (L3, L7) participated in both the co-design study and the user study. ~\ddag~ indicates progressive visual impairment. Education level abbreviations: GR (Graduate), UG (Undergraduate), HS (High School). \emph{FoV} abbreviates field of vision.}
    \label{tab:study_participants}
\end{table*}

\subsubsection{Materials}
\begin{figure}[htbp]
    \centering
    \includegraphics[width=\linewidth]{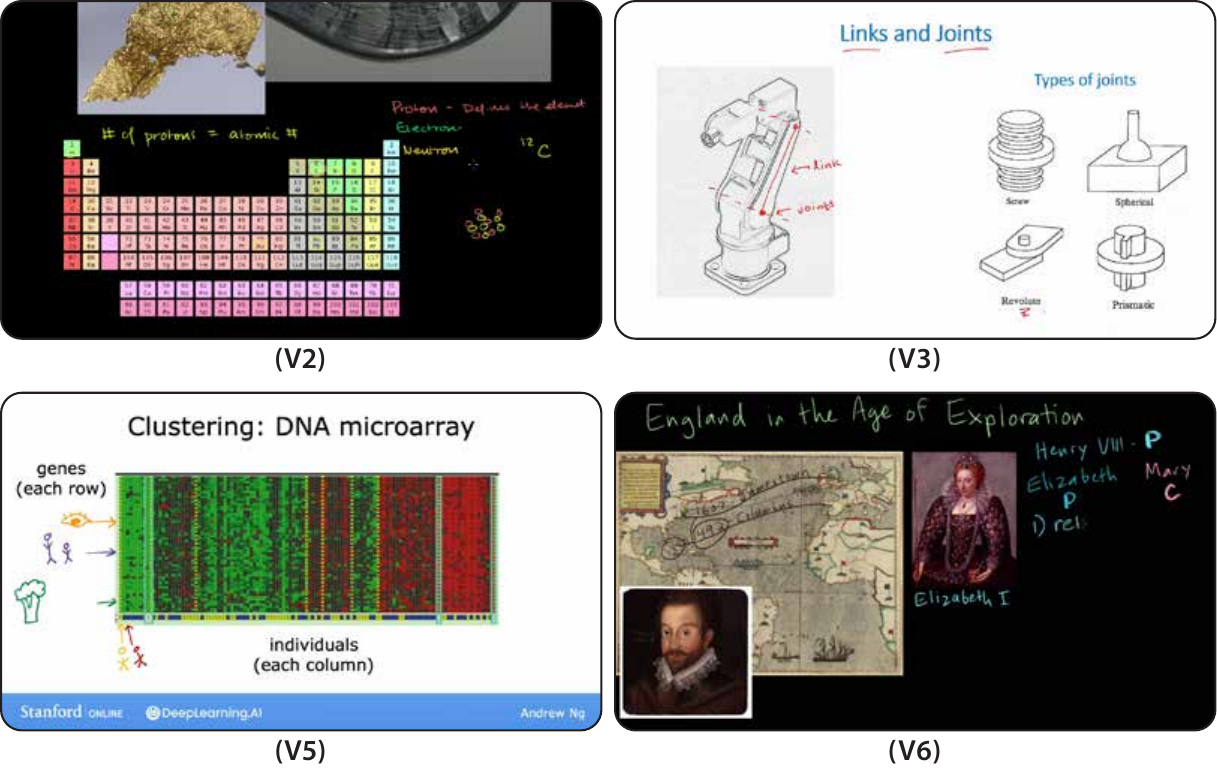}
    \caption{
        Sample frames from some of the videos used in the study, illustrating the visual diversity of the study videos. V2 and V6 are videos by Khan Academy, all Khan Academy content is available for free at ~\url{http://www.khanacademy.org/}. V3 is courtesy of ThatsEngineering (\url{https://www.youtube.com/@thatsengineeringsl}). V5 is courtesy of DeepLearning.AI (\url{https://www.deeplearning.ai/}).
    }
    \Description{
        Four frames from videos used in the user study. V2 (Chemistry): A scientific illustration with a mixed layout, this frame combines a microscopic image of gold, a periodic table in vibrant pastel colors, and molecular notations. The cursor is a small, thin crosshair, and sketch lines are thin, using multiple colors. V3 (Robotics): This grayscale sketch-style frame explains mechanical links and joints, featuring a robotic arm and annotated diagrams. It uses hand-drawn-like lines and text annotations in thin red strokes, with no pointer visible and no images, only blueprints on a bright, plain background. V5 (Computer Science): This frame features a white background showing a gene sequencing diagram with green and red colors and to the side several hand-draws sketches using multiple colors. No pointer is visible. V6 (World History): A warm-toned historical map of England, this frame is overlaid with labeled locations, portraits, handwritten-style annotations, and an old-fashioned visual theme. The cursor is subtle with a crosshair design, and there is an inset with a portrait.
    }
    \label{fig:study_videos_snapshot}
\end{figure}

In the study, we used a set of six educational videos sourced from YouTube~\cite{youtube}, DeepLearning.AI~\cite{deeplearningai}, and Khan Academy~\cite{khanacademy}: four short videos (1--2 minutes duration) and two longer videos (5--7 minutes duration), see Table~\ref{tab:study_videos}.
These videos were selected to represent a variety of academic disciplines, content creators, and visual presentation styles, aiming to approximate diverse real-world viewing scenarios.
An additional video, which was an extracted segment from the co-design study video (Section~\ref{sec:co_design}), was used for the initial tutorial and practice sessions but was not included in the final analysis.
All videos contained screen-shared slides accompanied by the instructor's voice, who interacted with the slides through pointing, marking, or sketching gestures.
Snapshots demonstrating visual characteristics of study videos are shown in Figure~\ref{fig:study_videos_snapshot}.

\begin{table*}[htbp]
    \centering
    \begin{tabular}{R{1.5cm}lllcccccc}
    \toprule
    VID & Domain           & Discipline       & Length & No. Slides & No. Activities & Point                    & Sketch                   & Mark                     \\
    \midrule
    V1~\cite{v1}  & Natural Sciences & Biology          & 1m4s   & 1          & 5              & \checkmark               & \checkmark               & \checkmark               \\
    V2~\cite{v2}  & Natural Sciences & Chemistry        & 1m2s   & 1          & 8              & \checkmark               & \checkmark               & \textthreequartersemdash \\
    V3~\cite{v3}  & Applied Sciences  & Robotics         & 1m3s   & 1          & 10             & \checkmark               & \checkmark               & \checkmark               \\
    V4~\cite{v4}  & Natural Sciences & Neuroethology    & 1m2s   & 1          & 7              & \checkmark               & \textthreequartersemdash & \textthreequartersemdash \\
    \midrule
    V5~\cite{v5}  & Formal Sciences  & Computer Science & 7m3s   & 4          & 34             & \textthreequartersemdash & \checkmark               & \checkmark               \\
    V6~\cite{v6}  & Humanities       & World History    & 5m27s  & 1          & 67             & \checkmark               & \checkmark               & \textthreequartersemdash \\
    \bottomrule
\end{tabular}
    \caption{Information on videos used in our study. V1--V4 were used for the Localization Task (Task 1), and V5--V6 were used for the Viewing Task (Task 2).}
    \label{tab:study_videos}
\end{table*}

\subsubsection{Procedure}
The study involved participants completing two sequential tasks: first, a \textbf{localization task} designed to address RQ1--RQ3, followed by a \textbf{viewing task} targeting RQ4--RQ5.
To ensure fairness and external validity, all participants could pause videos and use their usual accessibility aids---such as screen magnifiers---in both highlight and non-highlight conditions. This choice mirrors everyday scenarios and supports individual needs.
We randomized and counterbalanced both the task videos and their order within each task and the order of the two conditions. Finally, we normalized results to the number of activities to ensure comparability across conditions.

Before beginning each task, participants received an overview of the interface and completed a guided tutorial.
To minimize potential eye strain and fatigue, we scheduled predefined breaks into the study: 2-minute breaks between segments of a task and 5-minute breaks between tasks.
Following the completion of both tasks, we conducted a semi-structured interview with each participant to gather qualitative feedback and insights into their experience.

All study sessions were conducted remotely using the Zoom~\cite{zoom} video conferencing platform, with participants using their personal computers and monitors. Participants shared their screen, which lets us observe when they used assistive tools, such as screen magnifiers. Sessions were audio and video recorded, and subsequently transcribed for detailed analysis using OpenAI Whisper~\cite{whisper}.

\paragraph{Task 1: Localization Task:}
To address RQ1--RQ3, participants watched videos V1--V4 and pressed the \verb|Enter| key when they visually identified a relevant activity (pointing, marking, sketching). Each visual search task of a target activity was paired with a short sound notification, prompting participants to begin their search. Activities shorter than one second were excluded. Identification was based on participants’ subjective judgment of having located the cued item; they were not asked to indicate its exact location. There was no time limit, but failing to respond before the next cue was recorded as a missed detection.
In both baseline and VeasyGuide conditions, participants viewed videos linearly, with the option to pause or switch the view to full-screen. The VeasyGuide condition included visual highlights around target activities but excluded magnification. Highlight visuals followed system defaults (Table~\ref{tab:highlight_settings}).
We collected reaction times (cue to key press), missed detections, and subjective workload using the NASA-TLX questionnaire with a 5-point Likert scale.

\paragraph{Task 2: Viewing Task:}
To address RQ4--RQ5, participants watched videos V5--V6 and focused on the educational content. After each video, they completed a short comprehension quiz to encourage attentive viewing. In this task, the VeasyGuide condition included both highlighting and automatic zoom features.
The task had a fixed time limit set to \(x1.5\) of the video duration. Participants retained full control over playback (play, pause, seek) but could not revisit earlier video segments after moving on to the quiz.
We collected telemetry data capturing player interactions (e.g., settings configuration, pauses, seeking) and subjective feedback via a post-task experience questionnaire adapted from prior work~\cite{mo2022video, hwang2013concept}, using a 5-point Likert scale.

\subsubsection{Analysis}
\label{sec:analysis}
For quantitative results, since different videos are used, we performed an unpaired analysis. First, we visualized the results (Appendix~\ref{apdx:normality_checks}) then used the Shapiro Wilk test to check for normality. Results show that success rates follow a normal distribution (\(W = .946\), \(p = .671\)), whereas detection speeds violate this assumption (\(W = .772\), \(p = .014\)). Therefore, we used the Welch’s t-test for success rate and the Mann-Whitney U test for detection speed.
For subjective evaluation, significance testing was performed with the Wilcoxon Signed-Rank test.
We analyzed qualitative data using reflexive thematic analysis~\cite{braun2019reflecting}, which foregrounds the first author’s perspective as an individual with low vision as a critical lens. This iterative process involved familiarization with transcripts and observations, systematic coding of relevant excerpts, and the development, review, and refinement of thematic categories.

\subsection{Results}
\label{sec:results}
Our analysis reveals that VeasyGuide offers significant advantages in addressing the visual search problem for LV users (RQ1--RQ3) and enhances the experience of educational presentation videos for both LV users (RQ4) and sighted users (RQ5).

\subsubsection{Impact on Noticing Visual Activities for LV Users (RQ1)}
\label{sec:results-locating}
\begin{figure}[ht]
    \centering
    \includegraphics[width=0.8\linewidth]{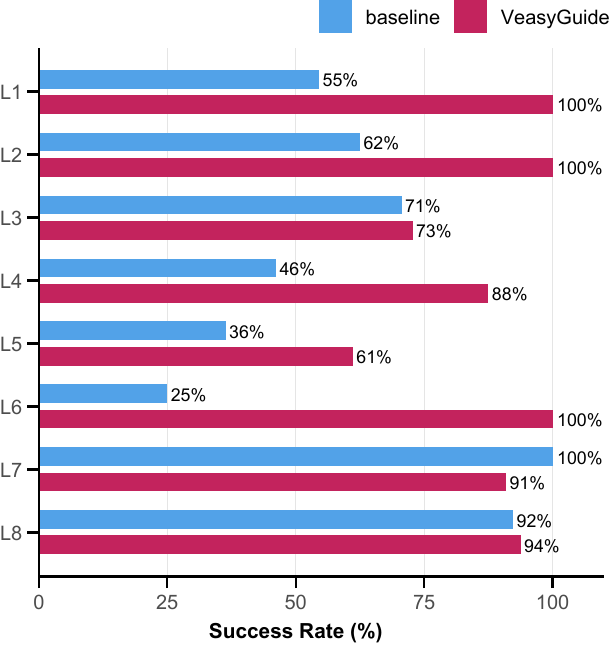}
    \caption{
        Visual search success rates of LV participants (L1-L8) in the localization task under baseline and VeasyGuide conditions.
    }
    \Description{
        Bar chart for success rates. Success rate is on the X-axis and is in percentage from 0 to 100. The Y-axis is the participant ID (L1-L8). There is a trend of improvement from baseline to VeasyGuide, with L1, L2, L4, L5, L6 showing the most improvement.
    }
    \label{fig:user_success_low_vision}
\end{figure}

We measured localization success rates as the proportion of correctly identified activities relative to the total number of cued activities presented in each condition's videos (Figure~\ref{fig:user_success_low_vision}).
For LV participants, a Welch’s t-test indicated that the difference was statistically significant, with VeasyGuide yielding higher success rates (means: baseline = \( 0.61 \), VeasyGuide = \( 0.88\); \(t(10.86) = -2.60\), \(p < 0.05\), \(\text{Cohen’s d} = 1.30\)). The mean rate of improvement in success rates between participants is \( 74.5\% \) (\( \pm 35\% \)) when comparing the baseline to VeasyGuide. See Appendix~\ref{apdx:full_user_results} for detailed breakdown of participants' results.

\begin{figure}[ht]
    \centering
    \includegraphics[width=0.8\linewidth]{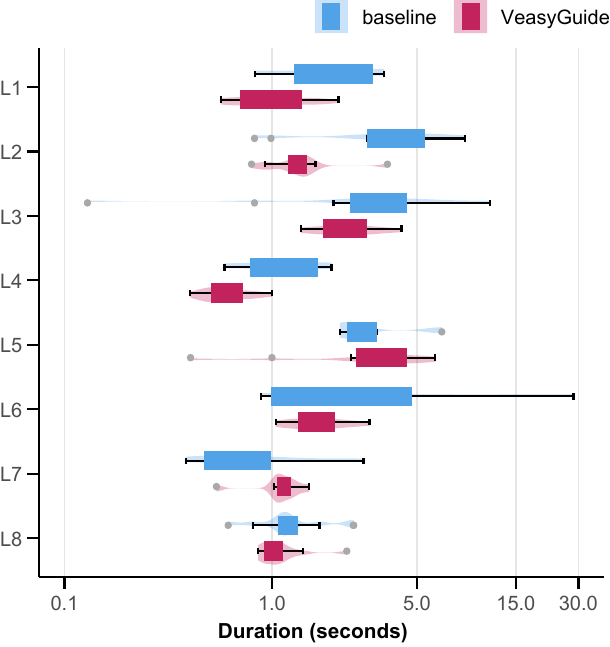}
    \caption{
        Visual search times of LV participants (L1-L8) in the localization task under baseline and VeasyGuide conditions.
    }
    \Description{
        Box plots for detection time on successful detection. Speed is on the X-axis and is in logarithmic scale from 0.1 to 30 seconds. The Y-axis is the participant ID (L1-L8). There is a trend of lower times for VeasyGuide.
    }
    \label{fig:user_speed_low_vision}
\end{figure}

\subsubsection{Impact on Visual Search Time for LV Users (RQ2)}
\label{sec:results-search-time}
Visual search time was defined as the duration from the moment an activity was presented until the participant pressed the key to indicate localization. This duration included any time the video might have been paused by the participant (Figure~\ref{fig:user_speed_low_vision}).
While we observe a general trend of improvement, a Mann-Whitney U test reveals no statistically significant difference. We attribute the observed results to differences in participants’ visual abilities, varying subjective video difficulty (Appendix~\ref{apdx:video_user_results}), and the default style, derived from the co-design study, which may not fully accommodate all participants’ needs.
The median search times for LV participants across all successful localizations were \( 1.38s\) (\( IQR = 2.03 \)) with the baseline and \( 1.19s \) (\( IQR = 0.84 \)) with VeasyGuide. Similarly, the mean search times were \( 2.57s \) (\(  \pm 3.7s\)) with the baseline and \( 1.53s \) (\( \pm 1.11s \)) with VeasyGuide. The mean rate of improvement in search times between participants is \( 33.1\% \) (\( \pm 12.7\% \)) when comparing the baseline to VeasyGuide.
A detailed breakdown of participants' results is available in Appendix ~\ref{apdx:full_user_results}.

\begin{figure*}[htbp]
    \centering
    \includegraphics[width=\textwidth]{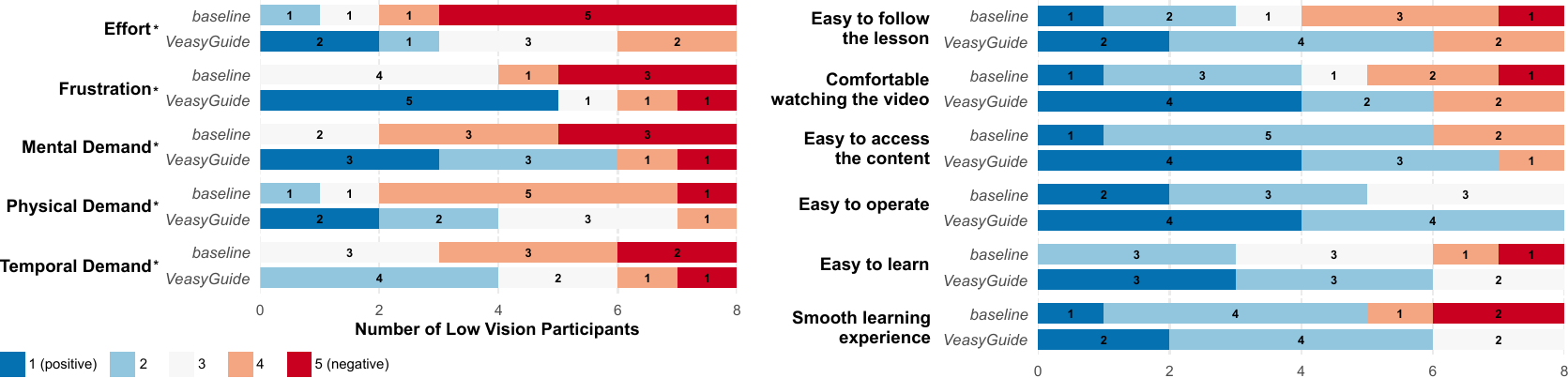}
    \caption{Results of subjective evaluation of LV participants comparing the baseline to VeasyGuide. Blue tones indicate positive evaluation and red tones indicate negative evaluation. We indicate statistical significance ($p < 0.05$) using \texttt{*}.}
    \Description{Horizontal stacked bar charts, where each bar represents a system's performance in a specific metric and the segments within the bar indicate the number of participants who selected each score. The data represents responses from participants with low vision. The scores range from 1 to 5, with 1 being the most positive and 5 being the most negative. Each bar is labeled with the metric (e.g., "Effort," "Frustration," "Easy to follow the lesson") and the system being evaluated ("baseline" or "VeasyGuide"). Not all bars are statistically significant, however all bars show a trend of improvement for VeasyGuide.}
    \label{fig:low_vision_subjective}
\end{figure*}

\subsubsection{Strategies for Visual Search (RQ3)}
\label{sec:results-strategies}
Our observations and participant feedback indicate that LV participants adopted diverse strategies to complete the localization task, with notable differences between the two conditions.
In the baseline condition, several participants frequently paused playback to gain time to locate cued activities (L2, L3, L5, L6). They also often used screen magnification (L1, L2, L5, L6), sometimes alongside pausing (L2, L6). For example, L6 repeatedly toggled pause and play to detect visual changes within a magnified view, while L2 further used a color inversion filter to enhance contrast when paused. Participants often scanned the magnified screen unsystematically, relying on guesswork to locate targets (L1, L2, L5). L6 was an exception, attempting a more systematic scan of the magnified frame. Despite these varied strategies, participants frequently reported difficulty or failure in locating activities.
In the VeasyGuide condition, no participants paused playback. Reliance on magnification decreased: only L1 and L5 used magnification, primarily to complement the visual highlights that guided and localized their search. For others, the highlights alone were sufficient for visual search without magnification.

\subsubsection{Impact on Accessibility of Presentations for LV Users (RQ4)}
\label{sec:results-accessibility}
We analyzed the impact of VeasyGuide on the accessibility and overall viewing experience of educational presentation videos for LV participants.
In the viewing task, participants watched a presentation video (V5-V6) and answered a quiz consisting of six questions. The mean quiz scores for baseline and VeasyGuide were \( 83.3\% \) (\( \pm 15.4\% \)) and \( 81.2\% \) (\( \pm 22.5\% \)), respectively; mean completion times were \( 2.28 \) (\( \pm 1.29 \)) and \( 2.39 \) (\( \pm 1.30 \)) minutes (Appendix~\ref{apdx:quiz_completion_time}). Statistical analysis indicated no significant difference.

\paragraph{Experience and Engagement}
A paired Wilcoxon Signed-Rank Test comparing NASA-TLX workload scores between the baseline and VeasyGuide conditions revealed statistically significant reductions across all assessed dimensions, all exhibiting large effect sizes (\( r > 0.5 \)), favoring VeasyGuide.
We observed significant reductions in \emph{Mental demand} (median difference = -1.5, \( V = 34.5 \), \( Z = 2.34 \), \( p < 0.05 \), \( r = 0.59 \)), \emph{Physical demand} (median difference = -1.0, \( V = 21 \), \( Z = 2.35 \), \( p < 0.05 \), \( r = 0.59 \)), \emph{Temporal demand} (median difference = -0.5, \( V = 21 \), \( Z = 2.37 \), \( p < 0.05 \), \( r = 0.59 \)), \emph{Effort} (median difference = -1.5, \( V = 28 \), \( Z = 2.48 \), \( p < 0.05 \), \( r = 0.62 \)), and \emph{Frustration} (median difference = -1.5, \( V = 21 \), \( Z = 2.35 \), \( p < 0.05 \), \( r = 0.59 \)).

Qualitative feedback aligned closely with the quantitative results. Participants consistently reported that VeasyGuide reduced the stress of tracking visual content. L1 noted that in the baseline condition, the difficulty of following visual elements might lead them to abandon the video in favor of reading a textbook. L5 highlighted this reduction in cognitive load: \textit{``I was relaxed listening to [the presentation] instead of constantly having to worry about where the next thing was going to appear.''}

Most participants who use a screen magnifier valued the zoom feature in VeasyGuide (L1, L2, L5), as it enabled them to follow highlighted activities without manually adjusting their screen magnifier. Others, such as L7, appreciated how the zoom helped them focus, \textit{``I found myself actually focusing, and it’s rare.''}
Moreover, L1 and L5 praised the automatic transitions between activities for reducing cognitive effort.
However, some participants (L1, L5, L6) requested manual control to override or reposition the VeasyGuide zoom window.

\begin{figure*}[htbp]
    \centering
    \includegraphics[width=\linewidth]{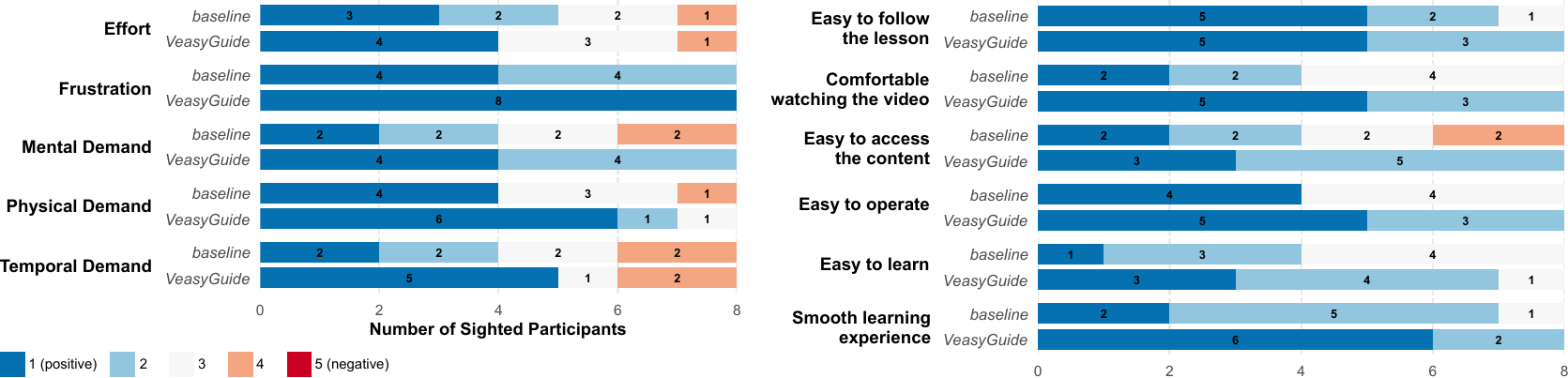}
    \caption{Results of subjective evaluation of sighted users comparing the baseline to VeasyGuide. Blue tones indicate positive evaluation and red tones indicate negative evaluation.}
    \Description{Horizontal stacked bar charts, where each bar represents a system's performance in a specific metric and the segments within the bar indicate the number of participants who selected each score. The data represents responses from participants who are sighted. The scores range from 1 to 5, with 1 being the most positive and 5 being the most negative. Each bar is labeled with the metric (e.g., "Effort," "Frustration," "Easy to follow the lesson") and the system being evaluated ("baseline" or "VeasyGuide"). No bar is indicated as statistically significant, however all bars show minor trend of improvement for VeasyGuide.}
    \label{fig:sighted_subjective}
\end{figure*}

\paragraph{Personalization and Perceived Accessibility}
Based on the experience questionnaire, while a paired Wilcoxon Signed-Rank Test revealed no significant difference, LV participants rated VeasyGuide more positively than the baseline across all dimensions (Figure~\ref{fig:low_vision_subjective}).
Most LV participants actively personalized VeasyGuide, often adjusting highlight styles (see Appendix~\ref{apdx:personalization_usage}).  Most commonly, participants modified the fill and border styles.  Participants with a limited field of vision (L1, L3) enabled highlight animation. Most participants found the optional accompanying pointer helpful; only L1 and L2 chose to disable it. Several participants (L5, L6, L7) applied post-processing filters---particularly color inversion---to reduce eye strain (L5) or improve the contrast of highlights against the video (L5, L6). Participants valued the ability to make these adjustments in real-time. As L2, an accessibility expert, noted, \textit{``[Personalization is] very helpful for people that can experience fluctuating vision loss.''}

Participants also described an enhanced sense of connection and engagement with the video instructor, attributing this to the guidance provided by the highlights. L3, who has a narrow field of vision, remarked that the highlights helped maintain focus on the instructor's points of emphasis. L1 articulated a core accessibility challenge addressed by the tool: \textit{``One of the biggest problems is that I don't know what in the video requires my attention. So the highlight system goes a long way to address the problem. It offloads this difficulty to software. I strain less.''}
Furthermore, some participants (L1, L2, L3) felt the highlights effectively conveyed the instructor’s focus and intent, strengthening their connection to the lecture content. L3 elaborated on this point: \textit{``It really helped me to focus better on the video, it helped me stay with the instructor. [...] Thanks to the animation, I saw that [the instructor] was drawing a boat, and I was happy I did not miss it.''}

Despite the predominantly positive feedback, some participants (L1, L4, L7, L8) acknowledged a learning curve associated with adopting a new tool like VeasyGuide. L1 reflected, \textit{``Getting used to it and figuring out my own workflow would let me worry less about the visual part and focus more on the content.''} Similarly, L8 felt more time would be needed to fully integrate VeasyGuide into their viewing habits. Additionally, several participants suggested that the benefits of VeasyGuide might be particularly pronounced in STEM\footnote{Science, Technology, Engineering, and Mathematics} subjects, which often feature dense visual information. L7 specifically mentioned mathematics lectures as a context where VeasyGuide could offer substantial advantages. In this study we selected presentation videos with a broader appeal, however, future research should investigate its effectiveness within STEM subjects.

\subsubsection{Impact of VeasyGuide on Sighted Users (RQ5)}
\label{sec:results-sighted}
Though built for LV users, to assess broader applicability, we also evaluated VeasyGuide with sighted participants. Objective metrics showed no significant differences (success: $\text{median}=1$, $\text{IQR}=0$; speed: $1.05\,\text{s}$ vs.\ $0.95\,\text{s}$), but subjective feedback favored VeasyGuide (Figure~\ref{fig:sighted_subjective}).

Sighted participants reported that VeasyGuide’s visual highlights and magnification improved their focus and comprehension. Participants noted that the highlights helped them maintain attention on relevant visual elements (S1--S4, S7, S8), reduced perceived visual clutter (S4--S6), and clarified the instructor’s intent (S2, S8). For example, S1 stated, \textit{``If I’m bored, the highlights help draw my attention back to the video; and to the right place.''} Similarly, S7 remarked, \textit{``It was easier to know what was important by following the highlight.''} Reflecting on the experience, S2 commented, \textit{``I could just focus on receiving information. It allowed me to focus on the learning part.''} Most sighted participants (S2--S5, S8) customized the highlight appearance, often based on aesthetic preferences. For instance, S2 selected a pink highlight simply because they liked the color.

Five sighted participants utilized the magnification feature (S1, S3, S5--S7). They reported that zooming helped direct their visual attention and minimize distractions from surrounding content. S1 commented that zooming \textit{``synchronizes the learner and teacher on the same section of the same [metaphoric] page.''} Others noted that zoom helped streamline the viewing environment by effectively clipping out peripheral content (S5--S7). S5 explained this preference: \textit{``if I zoomed out, I would have too many elements on the screen to manage.''} Finally, similar to LV participants, several sighted participants (S5--S8) expressed that VeasyGuide's benefits could be even more substantial in content-dense STEM fields.

Collectively, these findings suggest that the highlights and zoom features of VeasyGuide, which are designed primarily for accessibility, can offer broader benefits related to clarity, attention management, and user control, potentially enhancing the learning experience for a wider audience.
\section{Discussion}
\label{sec:discussion}

Our findings demonstrate that VeasyGuide enhances low-vision (LV) users' ability to detect visual activities, and reduces the visual burden and cognitive effort needed (Sections~\ref{sec:results-locating}--\ref{sec:results-accessibility}). Although speed improvements in activity detection varied, the success of LV participants in noticing visual activities increased significantly, with decreased reliance on magnification for detecting activities. The preference for the box-shaped highlight affirms the value of conveying spatial information, echoing insights from prior work~\cite{Jiang2024Context}.
Furthermore, VeasyGuide improved the learning experience, enabling both LV and sighted users to concentrate and engage more with the content. For sighted participants, the zoom feature, while not needed for visual clarity, reduced distractions by narrowing the visible content (Section ~\ref{sec:results-sighted}).

The remainder of this section discusses broader implications of our findings. We outline design principles for technologies addressing visual search challenges for LV users (Section~\ref{sec:design_space}), explore the applicability of our approach to other domains (Section~\ref{sec:applications}), acknowledge study limitations, and propose future research directions (Section~\ref{sec:limitations_future}).

\subsection{Design Framework for the Visual Search Problem}
\label{sec:design_space}
Visual search poses an ongoing challenge (\textbf{DI1}) for individuals with LV~\cite{wang2024low,JiCZ22,zhao2016cuesee}, primarily due to difficulties in determining \emph{what} to look for and \emph{where} to look.
While many LV users prefer leveraging their residual vision~\cite{szpiro2016people,zhao2016cuesee}, most current tools rely on audio description (AD) to convey visual content~\cite{Fan2023ImprovingAccessibility,stangl2023potential,ning2024spica}.
Building on the design implications (DIs) we outline in Section~\ref{sec:design_implications} and insights from our user study, we introduce a framework for designing tools that support LV users in performing visual search tasks.

\subsubsection{Familiarity with target visuals}
Familiarity with target visuals (\textbf{DI2}) helps LV users identify \emph{what} to look for. Co-design participants noted that inconsistent or unfamiliar video visuals (e.g., color choices, pointer styles) made it difficult to notice instructor actions. VeasyGuide fostered familiarity through three strategies: consistent highlight styles, recognizable pointer icons, and personalization. While we hypothesized that type-specific highlights would improve predictability, participants reported that consistent styles were more effective for noticing activity. Recognizable pointer icons, co-designed to resemble a computer cursor, further enhanced familiarity; some participants even wanted to match it to their own. During the evaluation, most LV participants enabled this indicator, with only two disabling it. Allowing users to personalize highlight visuals also increased engagement and reinforced their mental model of what to focus on. This aligns with findings from \citet{cunningham2014self}, who demonstrated that even young children exhibit enhanced memory for objects associated with themselves. Personalization allows users to connect highlights with their own needs through self-representation.

\subsubsection{Predictability of target spatial context}
Predictability (\textbf{DI3}) supports LV users in determining \emph{where} to look. Although spatial context is meaningful~\cite{hinojosa2015investigations}, it is often inaccessible to LV users~\cite{d2021accessible,Jiang2024Context}. Prior work conveys location using audio~\cite{d2021accessible,Fan2023ImprovingAccessibility}, visual~\cite{fox2023using,zhao2016cuesee,billah18SteeringWheel}, and tactile modalities~\cite{d2021accessible}.
Our findings show that, beyond location, conveying expected trajectory is critical for supporting visual search. VeasyGuide introduced a box-shaped highlight that encoded both spatial location and gesture trajectory using the box's width and height. For example, a short, narrow highlight indicated an underlining gesture, where the direction was inferred from the pen's position and the box's shape. In our evaluation, all but one participant used the box-shaped highlight.

Furthermore, based on participant feedback, we implemented a short pre-activity activation phase before the highlight was fully rendered. For instance, participants could anticipate the location of upcoming actions, such as sketching. This feature allowed users to adjust their gaze in advance, improving comfort and reducing temporal pressure during the activity. By providing both spatial location and motion trajectory, VeasyGuide narrowed the visual search space and improved anticipability. This helped reduce stress and scanning effort, particularly for screen magnifier users.

\subsubsection{Personalization with instant visual feedback}
Beyond familiarity, personalization (\textbf{DI4}) enabled users to configure highlight styles across various video formats (e.g., slide-based presentations or screen recordings), accommodate fluctuating or situational vision, and tailor the interface to individual needs. LV users have diverse and dynamic requirements~\cite{leat1999low,zhao2020Designing,wang2023understanding}, influenced by both vision conditions and environmental factors. These needs can be in tension; for example, animations may benefit users with a narrow field of vision but distract those with photosensitivity. Prior work in augmented reality~\cite{zhao2020Designing} has shown that preferences vary even among users with similar visual profiles.
While previous studies emphasize the importance of flexibility, our findings highlight the benefits of making personalization responsive, and integrate it into the interface. VeasyGuide introduces instant visual feedback as a core feature, allowing users to see the effects of customization in real-time.

In our study, most participants adjusted highlight styles in different ways (Section~\ref{sec:results-accessibility}), and instant feedback reduced the cognitive effort involved in selecting appropriate styles. This approach helps users respond to the unpredictable nature of visual search, which varies across content, context, and time. However, many personalization options may overwhelm users, leading to decision fatigue and a steeper learning curve. Future systems might mitigate this by offering user-managed defaults~\cite{sackl2020ensuring} or simplified contextual presets~\cite{zhao2020Designing} (e.g., photophobia, visual acuity, narrow field of vision).

\subsubsection{Support for LV user agency}
While DI2--DI4 address ambiguity in what and where to look, our findings reveal an additional need: preserving LV user agency during visual search. VeasyGuide enhances agency by giving users control over their visual experience through visible highlights. These highlights not only direct attention to instructor actions but also allow users to decide what content to attend to or ignore---choices which exist for sighted users. Visual search in educational videos can be cognitively demanding for LV users. By offering timely and visible cues, VeasyGuide reduces search effort and narrows the gap between the experience of sighted and LV users. With detection success gap reduced by \( 80.5\% \), and detection speed gap reduced by \( 27.3\% \). Rather than replacing vision, it reinforces residual vision use and supports independent learning strategies. Some participants expressed interest in even more control, such as adjusting the magnification window or navigating between recent highlights. Future systems should continue to preserve user agency and support users' preference to use their residual vision. This extends existing DIs by highlighting how users engage with visual content on their own terms, reinforcing agency as a foundation for effective visual search support.

\subsection{Visual Access in Other Domains}
\label{sec:applications}
VeasyGuide uses visual highlights to guide LV users in locating areas of interest during visual search tasks. While we focused on instructor actions---such as pointing, marking, and sketching---in educational videos, visual search problems also arise in other domains. Everyday scenarios offer opportunities to support visual search, including product identification while shopping~\cite{zhao2016cuesee}, and obstacle avoidance~\cite{fox2023using}.
VeasyGuide can be seamlessly integrated into mainstream platforms such as YouTube~\cite{youtube}, Coursera~\cite{coursera}, or Khan Academy~\cite{khanacademy}, benefiting LV learners on educational presentation videos.
Although live content, such as online lectures, represents a natural extension of VeasyGuide, the lack of complete spatial context during visual activities presents a significant challenge. VeasyGuide could address this issue by dynamically expanding highlights as additional activity information becomes available. Furthermore, because VeasyGuide employs computationally lightweight techniques, processing can be done on-device, which is needed for adaptation to live content.

In narrative media such as movies, preserving immersion should guide the design of assistive tools for visual search. Although AD is a common assistive approach, its additive nature can disrupt immersion. Leveraging narrative techniques, such as speed adjustments, camera movements, or zoom modifications may better support visual search without compromising the immersive experience.
Although visual highlights, such as those implemented in VeasyGuide, effectively support visual search, they may disrupt the immersive experience. In contrast, more subtle diegetic highlights, such as gentle brightening of segmented relevant on-screen elements or smooth color modifications tailored through personalization—could better balance visual clarity with immersion. For example, in some video games, creators use a specific color to direct players' attention and signify game elements they should interact with~\cite{Gamerant_NaughtyDog_Yellow}.

\section{Limitations and Future Work}
\label{sec:limitations_future}

This study has several limitations. First, the small sample size may not capture the full diversity of visual abilities among LV users. Moreover, as discussed in Section~\ref{sec:co-design_reflections}, VeasyGuide introduces a visual-access approach that was unfamiliar to some participants. Therefore, the short study duration may not have given them enough time to adapt, especially when the tool conflicted with their existing strategies for accessing educational video content~\cite{szpiro2016people}.

In this work, we did not evaluate the effect of personalization and familiarity in the Localization Task, instead fixing the settings to the default configuration (Section ~\ref{sec:results-search-time}). Such an evaluation would have required frequent adjustments across varying video styles, potentially increasing eye strain, fatigue, and cognitive load (see Section~\ref{sec:co-design_reflections}). Instead, we relied on a qualitative assessment of personalization and familiarity in the Viewing Task (Section~\ref{sec:results-accessibility}). Future work should devise methods that systematically study personalization while minimizing participant burden.

By encouraging LV users to engage with more visual content they might otherwise disregard, VeasyGuide may introduce unforeseen visual fatigue, which is an area that warrants further investigation.
Larger and more diverse longitudinal studies are needed to uncover long-term benefits, identify emerging interaction patterns, and assess potential visual fatigue.
Further, evaluating VeasyGuide in naturalistic educational settings could yield insights into its scalability and ecological validity.

Since VeasyGuide currently relies on context-free activity detection~\cite{parveen2021motion}, future research could add context-based cues. For example, during pointing actions, slide elements could be highlighted instead of the pointer itself, and during sketching, strokes could be previewed before they appear---similar to \citet{baudisch2006phosphor}---to convey location and trajectory more effectively.
Participants also suggested activity-based video navigation, echoing interaction modes in prior work~\cite{monserrat2013notevideo}. Exploring such interaction methods is another promising direction for future iterations of VeasyGuide.

\section{Conclusion}
\label{sec:conclusion}
In presentation videos, instructors frequently use digital pointers or pens to highlight, mark, or sketch content, providing visual cues intended to support learning. However, low-vision (LV) learners often miss these subtle visual actions, which can result in missed information and increased cognitive effort. Through a co-design study with LV participants, we identified barriers to visual search in instructional videos and developed VeasyGuide, an accessibility tool that detects and visualizes these cues using real-time highlights and optional zoom. VeasyGuide offers extensive personalization options, allowing users to tailor visualizations to their individual preferences and visual requirements. Our user study with LV participants demonstrated that VeasyGuide improves detection of visual activities and reduces cognitive workload. An additional evaluation with sighted participants indicates that VeasyGuide can enhance focus, comprehension, and engagement in video-based learning for both LV and sighted users. Drawing on our findings, we outlined a design space for supporting LV users in visual search tasks, emphasizing visual familiarity, predictable spatial context, personalization, and user agency. We believe that VeasyGuide can advance accessible media design and foster more equitable video learning experiences for all learners.

\begin{acks}
    We are grateful to the anonymous reviewers for their insightful feedback, and to all participants whose contributions enriched this study. This work was supported by JST AdCORP, Grant Number JPMJKB2302, Japan. Our gratitude extends to Google for endorsing this work through the Google Award for Inclusion Research (AIR) Grant. 
\end{acks}

\bibliographystyle{ACM-Reference-Format}
\bibliography{ref.main, ref.global}


\begin{thebibliography}{89}


\ifx \showCODEN    \undefined \def \showCODEN     #1{\unskip}     \fi
\ifx \showISBNx    \undefined \def \showISBNx     #1{\unskip}     \fi
\ifx \showISBNxiii \undefined \def \showISBNxiii  #1{\unskip}     \fi
\ifx \showISSN     \undefined \def \showISSN      #1{\unskip}     \fi
\ifx \showLCCN     \undefined \def \showLCCN      #1{\unskip}     \fi
\ifx \shownote     \undefined \def \shownote      #1{#1}          \fi
\ifx \showarticletitle \undefined \def \showarticletitle #1{#1}   \fi
\ifx \showURL      \undefined \def \showURL       {\relax}        \fi
\providecommand\bibfield[2]{#2}
\providecommand\bibinfo[2]{#2}
\providecommand\natexlab[1]{#1}
\providecommand\showeprint[2][]{arXiv:#2}

\bibitem[Academy(2024a)]%
        {v2}
\bibfield{author}{\bibinfo{person}{Khan Academy}.} \bibinfo{year}{2024}\natexlab{a}.
\newblock \bibinfo{title}{Elements and atoms | Atoms, compounds, and ions | Chemistry | Khan Academy}.
\newblock \bibinfo{howpublished}{\url{https://www.youtube.com/watch?v=IFKnq9QM6_A}}.
\newblock
\newblock
\shownote{Accessed: 2024-09-12}.


\bibitem[Academy(2024b)]%
        {v6}
\bibfield{author}{\bibinfo{person}{Khan Academy}.} \bibinfo{year}{2024}\natexlab{b}.
\newblock \bibinfo{title}{England in the Age of Exploration}.
\newblock \bibinfo{howpublished}{\url{https://www.youtube.com/watch?v=DZxG2miqgZI}}.
\newblock
\newblock
\shownote{Accessed: 2024-09-12}.


\bibitem[Aljedaani et~al\mbox{.}(2023)]%
        {wajdi2023blackboardApp}
\bibfield{author}{\bibinfo{person}{Wajdi Aljedaani}, \bibinfo{person}{Mohammed Alkahtani}, \bibinfo{person}{Stephanie Ludi}, \bibinfo{person}{Mohamed~Wiem Mkaouer}, \bibinfo{person}{Marcelo~M. Eler}, \bibinfo{person}{Marouane Kessentini}, {and} \bibinfo{person}{Ali Ouni}.} \bibinfo{year}{2023}\natexlab{}.
\newblock \showarticletitle{The State of Accessibility in Blackboard: Survey and User Reviews Case Study}. In \bibinfo{booktitle}{\emph{Proceedings of the 20th International Web for All Conference}} (<conf-loc>, <city>Austin</city>, <state>TX</state>, <country>USA</country>, </conf-loc>) \emph{(\bibinfo{series}{W4A '23})}. \bibinfo{publisher}{Association for Computing Machinery}, \bibinfo{address}{New York, NY, USA}, \bibinfo{pages}{84–95}.
\newblock
\showISBNx{9798400707483}
\href{https://doi.org/10.1145/3587281.3587291}{doi:\nolinkurl{10.1145/3587281.3587291}}


\bibitem[Aydin et~al\mbox{.}(2020)]%
        {aydin2020videoMagnif}
\bibfield{author}{\bibinfo{person}{Ali~Selman Aydin}, \bibinfo{person}{Shirin Feiz}, \bibinfo{person}{Vikas Ashok}, {and} \bibinfo{person}{IV Ramakrishnan}.} \bibinfo{year}{2020}\natexlab{}.
\newblock \showarticletitle{Towards making videos accessible for low vision screen magnifier users}. In \bibinfo{booktitle}{\emph{Proceedings of the 25th International Conference on Intelligent User Interfaces}} (Cagliari, Italy) \emph{(\bibinfo{series}{IUI '20})}. \bibinfo{publisher}{Association for Computing Machinery}, \bibinfo{address}{New York, NY, USA}, \bibinfo{pages}{10–21}.
\newblock
\showISBNx{9781450371186}
\href{https://doi.org/10.1145/3377325.3377494}{doi:\nolinkurl{10.1145/3377325.3377494}}


\bibitem[Baudisch et~al\mbox{.}(2006)]%
        {baudisch2006phosphor}
\bibfield{author}{\bibinfo{person}{Patrick Baudisch}, \bibinfo{person}{Desney Tan}, \bibinfo{person}{Maxime Collomb}, \bibinfo{person}{Dan Robbins}, \bibinfo{person}{Ken Hinckley}, \bibinfo{person}{Maneesh Agrawala}, \bibinfo{person}{Shengdong Zhao}, {and} \bibinfo{person}{Gonzalo Ramos}.} \bibinfo{year}{2006}\natexlab{}.
\newblock \showarticletitle{Phosphor: explaining transitions in the user interface using afterglow effects}. In \bibinfo{booktitle}{\emph{Proceedings of the 19th annual ACM symposium on User interface software and technology}}. \bibinfo{publisher}{ACM}, \bibinfo{address}{New York, NY, USA}, \bibinfo{pages}{169--178}.
\newblock


\bibitem[Billah et~al\mbox{.}(2018)]%
        {billah18SteeringWheel}
\bibfield{author}{\bibinfo{person}{Syed~Masum Billah}, \bibinfo{person}{Vikas Ashok}, \bibinfo{person}{Donald~E. Porter}, {and} \bibinfo{person}{I.~V. Ramakrishnan}.} \bibinfo{year}{2018}\natexlab{}.
\newblock \showarticletitle{SteeringWheel: {A} Locality-Preserving Magnification Interface for Low Vision Web Browsing}. In \bibinfo{booktitle}{\emph{Proceedings of the 2018 {CHI} Conference on Human Factors in Computing Systems, {CHI} 2018, Montreal, QC, Canada, April 21-26, 2018}}, \bibfield{editor}{\bibinfo{person}{Regan~L. Mandryk}, \bibinfo{person}{Mark Hancock}, \bibinfo{person}{Mark Perry}, {and} \bibinfo{person}{Anna~L. Cox}} (Eds.). \bibinfo{publisher}{{ACM}}, \bibinfo{pages}{20}.
\newblock
\href{https://doi.org/10.1145/3173574.3173594}{doi:\nolinkurl{10.1145/3173574.3173594}}


\bibitem[Bognar(2022)]%
        {Gamerant_NaughtyDog_Yellow}
\bibfield{author}{\bibinfo{person}{Christian Bognar}.} \bibinfo{year}{2022}\natexlab{}.
\newblock \bibinfo{booktitle}{\emph{Naughty Dog's Obsession With Yellow Explained}}.
\newblock
\urldef\tempurl%
\url{https://gamerant.com/naughty-dog-yellow-color-coding-environments-progression-design/}
\showURL{%
\tempurl}
\newblock
\shownote{Accessed: 2025-04-17}.


\bibitem[Braun and Clarke(2019)]%
        {braun2019reflecting}
\bibfield{author}{\bibinfo{person}{Virginia Braun} {and} \bibinfo{person}{Victoria Clarke}.} \bibinfo{year}{2019}\natexlab{}.
\newblock \showarticletitle{Reflecting on reflexive thematic analysis}.
\newblock \bibinfo{journal}{\emph{Qualitative research in sport, exercise and health}} \bibinfo{volume}{11}, \bibinfo{number}{4} (\bibinfo{year}{2019}), \bibinfo{pages}{589--597}.
\newblock


\bibitem[Burgstahler and Cory(2010)]%
        {burgstahler2010universal}
\bibfield{author}{\bibinfo{person}{Sheryl~E Burgstahler} {and} \bibinfo{person}{Rebecca~C Cory}.} \bibinfo{year}{2010}\natexlab{}.
\newblock \bibinfo{booktitle}{\emph{Universal design in higher education: From principles to practice}}.
\newblock \bibinfo{publisher}{Harvard Education Press}.
\newblock


\bibitem[Campos et~al\mbox{.}(2023)]%
        {campos2023machine}
\bibfield{author}{\bibinfo{person}{Virg{\'\i}nia~P Campos}, \bibinfo{person}{Luiz~MG Gon{\c{c}}alves}, \bibinfo{person}{Wesnydy~L Ribeiro}, \bibinfo{person}{Tiago~MU Ara{\'u}jo}, \bibinfo{person}{Tha{\'\i}s~G Do~Rego}, \bibinfo{person}{Pedro~HV Figueiredo}, \bibinfo{person}{Suanny~FS Vieira}, \bibinfo{person}{Thiago~FS Costa}, \bibinfo{person}{Caio~C Moraes}, \bibinfo{person}{Alexandre~CS Cruz}, {et~al\mbox{.}}} \bibinfo{year}{2023}\natexlab{}.
\newblock \showarticletitle{Machine generation of audio description for blind and visually impaired people}.
\newblock \bibinfo{journal}{\emph{ACM Transactions on Accessible Computing}} \bibinfo{volume}{16}, \bibinfo{number}{2} (\bibinfo{year}{2023}), \bibinfo{pages}{1--28}.
\newblock


\bibitem[Chang et~al\mbox{.}(2024)]%
        {chang2024worldscribe}
\bibfield{author}{\bibinfo{person}{Ruei-Che Chang}, \bibinfo{person}{Yuxuan Liu}, {and} \bibinfo{person}{Anhong Guo}.} \bibinfo{year}{2024}\natexlab{}.
\newblock \showarticletitle{WorldScribe: Towards Context-Aware Live Visual Descriptions}. In \bibinfo{booktitle}{\emph{Proceedings of the 37th Annual ACM Symposium on User Interface Software and Technology}}. \bibinfo{pages}{1--18}.
\newblock


\bibitem[{Coursera}({[n.\,d.]})]%
        {coursera}
\bibfield{author}{\bibinfo{person}{{Coursera}}.} \bibinfo{year}{[n.\,d.]}\natexlab{}.
\newblock \bibinfo{title}{Coursera}.
\newblock \bibinfo{howpublished}{\url{https://www.coursera.org/}}.
\newblock
\newblock
\shownote{Accessed: 2023-10-05}.


\bibitem[Cunningham et~al\mbox{.}(2014)]%
        {cunningham2014self}
\bibfield{author}{\bibinfo{person}{Sheila~J. Cunningham}, \bibinfo{person}{Joanne~L. Brebner}, \bibinfo{person}{Francis Quinn}, {and} \bibinfo{person}{David~J. Turk}.} \bibinfo{year}{2014}\natexlab{}.
\newblock \showarticletitle{The Self-Reference Effect on Memory in Early Childhood}.
\newblock \bibinfo{journal}{\emph{Child Development}} \bibinfo{volume}{85}, \bibinfo{number}{2} (\bibinfo{year}{2014}), \bibinfo{pages}{808--823}.
\newblock
\href{https://doi.org/10.1111/cdev.12144}{doi:\nolinkurl{10.1111/cdev.12144}}
\showeprint{https://srcd.onlinelibrary.wiley.com/doi/pdf/10.1111/cdev.12144}


\bibitem[D. and D.(2024)]%
        {pyscenedetect}
\bibfield{author}{\bibinfo{person}{Robert D.} {and} \bibinfo{person}{Adam D.}} \bibinfo{year}{2024}\natexlab{}.
\newblock \bibinfo{title}{PySceneDetect}.
\newblock
\urldef\tempurl%
\url{https://github.com/Breakthrough/PySceneDetect}
\showURL{%
\tempurl}


\bibitem[{DeepLearning.AI}({[n.\,d.]})]%
        {deeplearningai}
\bibfield{author}{\bibinfo{person}{{DeepLearning.AI}}.} \bibinfo{year}{[n.\,d.]}\natexlab{}.
\newblock \bibinfo{title}{DeepLearning.AI}.
\newblock \bibinfo{howpublished}{\url{https://www.deeplearning.ai/}}.
\newblock
\newblock
\shownote{Accessed: 2024-08-10}.


\bibitem[DeepLearningAI(2024)]%
        {v5}
\bibfield{author}{\bibinfo{person}{DeepLearningAI}.} \bibinfo{year}{2024}\natexlab{}.
\newblock \bibinfo{title}{\#6 Machine Learning Specialization [Course 1, Week 1, Lesson 2]}.
\newblock \bibinfo{howpublished}{\url{https://www.youtube.com/watch?v=gG_wI_uGfIE}}.
\newblock
\newblock
\shownote{Accessed: 2024-09-12}.


\bibitem[Denoue et~al\mbox{.}(2013)]%
        {denoue2013real}
\bibfield{author}{\bibinfo{person}{Laurent Denoue}, \bibinfo{person}{Scott Carter}, \bibinfo{person}{Matthew Cooper}, {and} \bibinfo{person}{John Adcock}.} \bibinfo{year}{2013}\natexlab{}.
\newblock \showarticletitle{Real-time direct manipulation of screen-based videos}. In \bibinfo{booktitle}{\emph{Proceedings of the companion publication of the 2013 international conference on Intelligent user interfaces companion}}. \bibinfo{pages}{43--44}.
\newblock


\bibitem[Developers(2024)]%
        {networkx}
\bibfield{author}{\bibinfo{person}{NetworkX Developers}.} \bibinfo{year}{2024}\natexlab{}.
\newblock \bibinfo{title}{NetworkX}.
\newblock
\urldef\tempurl%
\url{https://networkx.org/}
\showURL{%
\tempurl}


\bibitem[D’Agostino(2021)]%
        {d2021accessible}
\bibfield{author}{\bibinfo{person}{Alfred~T D’Agostino}.} \bibinfo{year}{2021}\natexlab{}.
\newblock \showarticletitle{Accessible teaching and learning in the undergraduate chemistry course and laboratory for blind and low-vision students}.
\newblock \bibinfo{journal}{\emph{Journal of Chemical Education}} \bibinfo{volume}{99}, \bibinfo{number}{1} (\bibinfo{year}{2021}), \bibinfo{pages}{140--147}.
\newblock


\bibitem[Elias et~al\mbox{.}(2018)]%
        {elias2018slidewiki}
\bibfield{author}{\bibinfo{person}{Mirette Elias}, \bibinfo{person}{Abi James}, \bibinfo{person}{Edna Ruckhaus}, \bibinfo{person}{Mari~Carmen Su{\'a}rez-Figueroa}, \bibinfo{person}{Klaas~Andries De~Graaf}, \bibinfo{person}{Ali Khalili}, \bibinfo{person}{Benjamin Wulff}, \bibinfo{person}{Steffen Lohmann}, {and} \bibinfo{person}{S{\"o}ren Auer}.} \bibinfo{year}{2018}\natexlab{}.
\newblock \showarticletitle{SlideWiki-Towards a Collaborative and Accessible Platform for Slide Presentations.}. In \bibinfo{booktitle}{\emph{EC-TEL (Practitioner Proceedings)}}. \bibinfo{pages}{1--3}.
\newblock


\bibitem[Facebook(2024)]%
        {react}
\bibfield{author}{\bibinfo{person}{Inc. Facebook}.} \bibinfo{year}{2024}\natexlab{}.
\newblock \bibinfo{title}{React - A JavaScript library for building user interfaces}.
\newblock \bibinfo{howpublished}{\url{https://reactjs.org}}.
\newblock
\newblock
\shownote{Accessed: 2024-09-12}.


\bibitem[Fan et~al\mbox{.}(2023)]%
        {Fan2023ImprovingAccessibility}
\bibfield{author}{\bibinfo{person}{Danyang Fan}, \bibinfo{person}{Sasa Junuzovic}, \bibinfo{person}{John~C. Tang}, {and} \bibinfo{person}{Thomas Jaeger}.} \bibinfo{year}{2023}\natexlab{}.
\newblock \showarticletitle{Improving the Accessibility of Screen-Shared Presentations by Enabling Concurrent Exploration}. In \bibinfo{booktitle}{\emph{Proceedings of the 25th International {ACM} {SIGACCESS} Conference on Computers and Accessibility, {ASSETS} 2023, New York, NY, USA, October 22-25, 2023}}. \bibinfo{publisher}{{ACM}}, \bibinfo{pages}{44:1--44:16}.
\newblock
\href{https://doi.org/10.1145/3597638.3608411}{doi:\nolinkurl{10.1145/3597638.3608411}}


\bibitem[Foundation(2024)]%
        {python}
\bibfield{author}{\bibinfo{person}{Python~Software Foundation}.} \bibinfo{year}{2024}\natexlab{}.
\newblock \bibinfo{title}{Python Programming Language}.
\newblock
\urldef\tempurl%
\url{https://www.python.org/}
\showURL{%
\tempurl}


\bibitem[Fox et~al\mbox{.}(2023)]%
        {fox2023using}
\bibfield{author}{\bibinfo{person}{Dylan~R. Fox}, \bibinfo{person}{Ahmad Ahmadzada}, \bibinfo{person}{Clara~T. Friedman}, \bibinfo{person}{Shiri Azenkot}, \bibinfo{person}{Marlena~A. Chu}, \bibinfo{person}{Roberto Manduchi}, {and} \bibinfo{person}{Emily~A. Cooper}.} \bibinfo{year}{2023}\natexlab{}.
\newblock \showarticletitle{Using augmented reality to cue obstacles for people with low vision}.
\newblock \bibinfo{journal}{\emph{Opt. Express}} \bibinfo{volume}{31}, \bibinfo{number}{4} (\bibinfo{date}{Feb} \bibinfo{year}{2023}), \bibinfo{pages}{6827--6848}.
\newblock
\href{https://doi.org/10.1364/OE.479258}{doi:\nolinkurl{10.1364/OE.479258}}


\bibitem[Ghosh and Figueroa(2023)]%
        {gosh2023tiktokLearning}
\bibfield{author}{\bibinfo{person}{Sourojit Ghosh} {and} \bibinfo{person}{Andrea Figueroa}.} \bibinfo{year}{2023}\natexlab{}.
\newblock \showarticletitle{Establishing TikTok as a Platform for Informal Learning: Evidence from Mixed-Methods Analysis of Creators and Viewers}. In \bibinfo{booktitle}{\emph{56th Hawaii International Conference on System Sciences, {HICSS} 2023, Maui, Hawaii, USA, January 3-6, 2023}}, \bibfield{editor}{\bibinfo{person}{Tung~X. Bui}} (Ed.). \bibinfo{publisher}{ScholarSpace}, \bibinfo{pages}{2431--2440}.
\newblock
\urldef\tempurl%
\url{https://hdl.handle.net/10125/102931}
\showURL{%
\tempurl}


\bibitem[Grice and Hughes(2009)]%
        {grice2009can}
\bibfield{author}{\bibinfo{person}{Sue Grice} {and} \bibinfo{person}{Janet Hughes}.} \bibinfo{year}{2009}\natexlab{}.
\newblock \showarticletitle{Can music and animation improve the flow and attainment in online learning?}
\newblock \bibinfo{journal}{\emph{Journal of Educational Multimedia and Hypermedia}} \bibinfo{volume}{18}, \bibinfo{number}{4} (\bibinfo{year}{2009}), \bibinfo{pages}{385--403}.
\newblock


\bibitem[Han et~al\mbox{.}(2023)]%
        {han2023autoad}
\bibfield{author}{\bibinfo{person}{Tengda Han}, \bibinfo{person}{Max Bain}, \bibinfo{person}{Arsha Nagrani}, \bibinfo{person}{G{\"u}l Varol}, \bibinfo{person}{Weidi Xie}, {and} \bibinfo{person}{Andrew Zisserman}.} \bibinfo{year}{2023}\natexlab{}.
\newblock \showarticletitle{AutoAD: Movie description in context}. In \bibinfo{booktitle}{\emph{Proceedings of the IEEE/CVF Conference on Computer Vision and Pattern Recognition}}. \bibinfo{pages}{18930--18940}.
\newblock


\bibitem[Hartsell and Yuen(2006)]%
        {hartsell2006video}
\bibfield{author}{\bibinfo{person}{Taralynn Hartsell} {and} \bibinfo{person}{Steve Chi-Yin Yuen}.} \bibinfo{year}{2006}\natexlab{}.
\newblock \showarticletitle{Video streaming in online learning}.
\newblock \bibinfo{journal}{\emph{AACE Review (Formerly AACE Journal)}} \bibinfo{volume}{14}, \bibinfo{number}{1} (\bibinfo{year}{2006}), \bibinfo{pages}{31--43}.
\newblock


\bibitem[Hinojosa(2015)]%
        {hinojosa2015investigations}
\bibfield{author}{\bibinfo{person}{Alfonso~Juan Hinojosa}.} \bibinfo{year}{2015}\natexlab{}.
\newblock \bibinfo{booktitle}{\emph{Investigations on the impact of spatial ability and scientific reasoning of student comprehension in physics, state assessment tests, and STEM courses}}.
\newblock \bibinfo{publisher}{The University of Texas at Arlington}, \bibinfo{address}{Arlington, TX, USA}.
\newblock


\bibitem[Hirvonen et~al\mbox{.}(2023)]%
        {hirvonen2023co}
\bibfield{author}{\bibinfo{person}{Maija Hirvonen}, \bibinfo{person}{Marika Hakola}, {and} \bibinfo{person}{Michael Klade}.} \bibinfo{year}{2023}\natexlab{}.
\newblock \showarticletitle{Co-translation, consultancy and joint authorship: User-centred translation and editing in collaborative audio description}.
\newblock \bibinfo{journal}{\emph{Journal of Specialised Translation}} \bibinfo{number}{39} (\bibinfo{year}{2023}), \bibinfo{pages}{26--51}.
\newblock


\bibitem[Hu(1962)]%
        {hu1962visual}
\bibfield{author}{\bibinfo{person}{Ming-Kuei Hu}.} \bibinfo{year}{1962}\natexlab{}.
\newblock \showarticletitle{Visual pattern recognition by moment invariants}.
\newblock \bibinfo{journal}{\emph{IRE transactions on information theory}} \bibinfo{volume}{8}, \bibinfo{number}{2} (\bibinfo{year}{1962}), \bibinfo{pages}{179--187}.
\newblock


\bibitem[Hwang et~al\mbox{.}(2013)]%
        {hwang2013concept}
\bibfield{author}{\bibinfo{person}{Gwo-Jen Hwang}, \bibinfo{person}{Li-Hsueh Yang}, {and} \bibinfo{person}{Sheng-Yuan Wang}.} \bibinfo{year}{2013}\natexlab{}.
\newblock \showarticletitle{A concept map-embedded educational computer game for improving students' learning performance in natural science courses}.
\newblock \bibinfo{journal}{\emph{Computers \& Education}}  \bibinfo{volume}{69} (\bibinfo{year}{2013}), \bibinfo{pages}{121--130}.
\newblock


\bibitem[Islam and Billah(2023)]%
        {islam23SpaceX}
\bibfield{author}{\bibinfo{person}{Md.~Touhidul Islam} {and} \bibinfo{person}{Syed~Masum Billah}.} \bibinfo{year}{2023}\natexlab{}.
\newblock \showarticletitle{SpaceX Mag: An Automatic, Scalable, and Rapid Space Compactor for Optimizing Smartphone App Interfaces for Low-Vision Users}.
\newblock \bibinfo{journal}{\emph{Proc. {ACM} Interact. Mob. Wearable Ubiquitous Technol.}} \bibinfo{volume}{7}, \bibinfo{number}{2} (\bibinfo{year}{2023}), \bibinfo{pages}{59:1--59:36}.
\newblock
\href{https://doi.org/10.1145/3596253}{doi:\nolinkurl{10.1145/3596253}}


\bibitem[Jacko and Sears(1998)]%
        {jacko1998designing}
\bibfield{author}{\bibinfo{person}{Julie~A Jacko} {and} \bibinfo{person}{Andrew Sears}.} \bibinfo{year}{1998}\natexlab{}.
\newblock \showarticletitle{Designing interfaces for an overlooked user group: Considering the visual profiles of partially sighted users}. In \bibinfo{booktitle}{\emph{Proceedings of the third international ACM conference on Assistive technologies}}. \bibinfo{pages}{75--77}.
\newblock


\bibitem[Jain et~al\mbox{.}(2023)]%
        {Jain2023Sports}
\bibfield{author}{\bibinfo{person}{Gaurav Jain}, \bibinfo{person}{Basel Hindi}, \bibinfo{person}{Connor Courtien}, \bibinfo{person}{Conrad Wyrick}, \bibinfo{person}{Xin Yi~Therese Xu}, \bibinfo{person}{Michael~C Malcolm}, {and} \bibinfo{person}{Brian~A. Smith}.} \bibinfo{year}{2023}\natexlab{}.
\newblock \showarticletitle{Towards Accessible Sports Broadcasts for Blind and Low-Vision Viewers}. In \bibinfo{booktitle}{\emph{Extended Abstracts of the 2023 CHI Conference on Human Factors in Computing Systems}} (Hamburg, Germany) \emph{(\bibinfo{series}{CHI EA '23})}. \bibinfo{publisher}{Association for Computing Machinery}, \bibinfo{address}{New York, NY, USA}, Article \bibinfo{articleno}{287}, \bibinfo{numpages}{7}~pages.
\newblock
\showISBNx{9781450394222}
\href{https://doi.org/10.1145/3544549.3585610}{doi:\nolinkurl{10.1145/3544549.3585610}}


\bibitem[Ji et~al\mbox{.}(2022)]%
        {JiCZ22}
\bibfield{author}{\bibinfo{person}{Tiger~F. Ji}, \bibinfo{person}{Brianna~R. Cochran}, {and} \bibinfo{person}{Yuhang Zhao}.} \bibinfo{year}{2022}\natexlab{}.
\newblock \showarticletitle{VRBubble: Enhancing Peripheral Awareness of Avatars for People with Visual Impairments in Social Virtual Reality}. In \bibinfo{booktitle}{\emph{Proceedings of the 24th International {ACM} {SIGACCESS} Conference on Computers and Accessibility, {ASSETS} 2022, Athens, Greece, October 23-26, 2022}}, \bibfield{editor}{\bibinfo{person}{Jon Froehlich}, \bibinfo{person}{Kristen Shinohara}, {and} \bibinfo{person}{Stephanie Ludi}} (Eds.). \bibinfo{publisher}{{ACM}}, \bibinfo{pages}{3:1--3:17}.
\newblock
\href{https://doi.org/10.1145/3517428.3544821}{doi:\nolinkurl{10.1145/3517428.3544821}}


\bibitem[Jiang et~al\mbox{.}(2024)]%
        {Jiang2024Context}
\bibfield{author}{\bibinfo{person}{Lucy Jiang}, \bibinfo{person}{Crescentia Jung}, \bibinfo{person}{Mahika Phutane}, \bibinfo{person}{Abigale Stangl}, {and} \bibinfo{person}{Shiri Azenkot}.} \bibinfo{year}{2024}\natexlab{}.
\newblock \showarticletitle{"It's Kind of Context Dependent": Understanding Blind and Low Vision People's Video Accessibility Preferences Across Viewing Scenarios}. In \bibinfo{booktitle}{\emph{Proceedings of the {CHI} Conference on Human Factors in Computing Systems, {CHI} 2024, Honolulu, HI, USA, May 11-16, 2024}}, \bibfield{editor}{\bibinfo{person}{Florian~'Floyd' Mueller}, \bibinfo{person}{Penny Kyburz}, \bibinfo{person}{Julie~R. Williamson}, \bibinfo{person}{Corina Sas}, \bibinfo{person}{Max~L. Wilson}, \bibinfo{person}{Phoebe O.~Toups Dugas}, {and} \bibinfo{person}{Irina Shklovski}} (Eds.). \bibinfo{publisher}{{ACM}}, \bibinfo{pages}{897:1--897:20}.
\newblock
\href{https://doi.org/10.1145/3613904.3642238}{doi:\nolinkurl{10.1145/3613904.3642238}}


\bibitem[Jiang and Ladner(2022)]%
        {jiang2022co}
\bibfield{author}{\bibinfo{person}{Lucy Jiang} {and} \bibinfo{person}{Richard Ladner}.} \bibinfo{year}{2022}\natexlab{}.
\newblock \showarticletitle{Co-Designing Systems to Support Blind and Low Vision Audio Description Writers}. In \bibinfo{booktitle}{\emph{Proceedings of the 24th International ACM SIGACCESS Conference on Computers and Accessibility}} (Athens, Greece) \emph{(\bibinfo{series}{ASSETS '22})}. \bibinfo{publisher}{Association for Computing Machinery}, \bibinfo{address}{New York, NY, USA}, Article \bibinfo{articleno}{74}, \bibinfo{numpages}{3}~pages.
\newblock
\showISBNx{9781450392587}
\href{https://doi.org/10.1145/3517428.3550394}{doi:\nolinkurl{10.1145/3517428.3550394}}


\bibitem[Jiang et~al\mbox{.}(2023)]%
        {jiang2023beyond}
\bibfield{author}{\bibinfo{person}{Lucy Jiang}, \bibinfo{person}{Mahika Phutane}, {and} \bibinfo{person}{Shiri Azenkot}.} \bibinfo{year}{2023}\natexlab{}.
\newblock \showarticletitle{Beyond audio description: Exploring 360 video accessibility with blind and low vision users through collaborative creation}. In \bibinfo{booktitle}{\emph{Proceedings of the 25th international ACM SIGACCESS conference on computers and accessibility}}. \bibinfo{pages}{1--17}.
\newblock


\bibitem[Jung et~al\mbox{.}(2018)]%
        {jung2018dynamicslide}
\bibfield{author}{\bibinfo{person}{Hyeungshik Jung}, \bibinfo{person}{Hijung~Valentina Shin}, {and} \bibinfo{person}{Juho Kim}.} \bibinfo{year}{2018}\natexlab{}.
\newblock \showarticletitle{Dynamicslide: Exploring the design space of reference-based interaction techniques for slide-based lecture videos}. In \bibinfo{booktitle}{\emph{Proceedings of the 2018 Workshop on Multimedia for Accessible Human Computer Interface}}. \bibinfo{publisher}{ACM}, \bibinfo{address}{New York, NY, USA}, \bibinfo{pages}{33--41}.
\newblock


\bibitem[{Khan Academy}({[n.\,d.]})]%
        {khanacademy}
\bibfield{author}{\bibinfo{person}{{Khan Academy}}.} \bibinfo{year}{[n.\,d.]}\natexlab{}.
\newblock \bibinfo{title}{Khan Academy}.
\newblock \bibinfo{howpublished}{\url{https://www.khanacademy.org/}}.
\newblock
\newblock
\shownote{Accessed: 2024-08-10}.


\bibitem[Kong et~al\mbox{.}(2021)]%
        {kong2021tutoriallens}
\bibfield{author}{\bibinfo{person}{Junhan Kong}, \bibinfo{person}{Dena Sabha}, \bibinfo{person}{Jeffrey~P Bigham}, \bibinfo{person}{Amy Pavel}, {and} \bibinfo{person}{Anhong Guo}.} \bibinfo{year}{2021}\natexlab{}.
\newblock \showarticletitle{TutorialLens: authoring Interactive augmented reality tutorials through narration and demonstration}. In \bibinfo{booktitle}{\emph{Proceedings of the 2021 ACM Symposium on Spatial User Interaction}}. \bibinfo{pages}{1--11}.
\newblock


\bibitem[Ladner and Rector(2017)]%
        {ladner2017making}
\bibfield{author}{\bibinfo{person}{Richard~E Ladner} {and} \bibinfo{person}{Kyle Rector}.} \bibinfo{year}{2017}\natexlab{}.
\newblock \showarticletitle{Making your presentation accessible}.
\newblock \bibinfo{journal}{\emph{Interactions}} \bibinfo{volume}{24}, \bibinfo{number}{4} (\bibinfo{year}{2017}), \bibinfo{pages}{56--59}.
\newblock


\bibitem[Leat et~al\mbox{.}(1999)]%
        {leat1999low}
\bibfield{author}{\bibinfo{person}{Susan~J Leat}, \bibinfo{person}{Gordon~E Legge}, {and} \bibinfo{person}{Mark~A Bullimore}.} \bibinfo{year}{1999}\natexlab{}.
\newblock \showarticletitle{What is low vision? A re-evaluation of definitions}.
\newblock \bibinfo{journal}{\emph{Optometry and Vision Science}} \bibinfo{volume}{76}, \bibinfo{number}{4} (\bibinfo{year}{1999}), \bibinfo{pages}{198--211}.
\newblock


\bibitem[Library(2024)]%
        {opencv}
\bibfield{author}{\bibinfo{person}{Open Source Computer~Vision Library}.} \bibinfo{year}{2024}\natexlab{}.
\newblock \bibinfo{title}{OpenCV}.
\newblock
\urldef\tempurl%
\url{https://opencv.org/}
\showURL{%
\tempurl}


\bibitem[Mattheiss et~al\mbox{.}(2017)]%
        {mattheiss2017user}
\bibfield{author}{\bibinfo{person}{Elke Mattheiss}, \bibinfo{person}{Georg Regal}, \bibinfo{person}{David Sellitsch}, {and} \bibinfo{person}{Manfred Tscheligi}.} \bibinfo{year}{2017}\natexlab{}.
\newblock \showarticletitle{User-centred design with visually impaired pupils: A case study of a game editor for orientation and mobility training}.
\newblock \bibinfo{journal}{\emph{International Journal of Child-Computer Interaction}}  \bibinfo{volume}{11} (\bibinfo{year}{2017}), \bibinfo{pages}{12--18}.
\newblock
\showISSN{2212-8689}
\href{https://doi.org/10.1016/j.ijcci.2016.11.001}{doi:\nolinkurl{10.1016/j.ijcci.2016.11.001}}
\newblock
\shownote{Designing with and for Children with Special Needs}.


\bibitem[Mo et~al\mbox{.}(2022)]%
        {mo2022video}
\bibfield{author}{\bibinfo{person}{Chuan-Yu Mo}, \bibinfo{person}{Chengliang Wang}, \bibinfo{person}{Jian Dai}, {and} \bibinfo{person}{Peiqi Jin}.} \bibinfo{year}{2022}\natexlab{}.
\newblock \showarticletitle{Video playback speed influence on learning effect from the perspective of personalized adaptive learning: A study based on cognitive load theory}.
\newblock \bibinfo{journal}{\emph{Frontiers in Psychology}}  \bibinfo{volume}{13} (\bibinfo{year}{2022}), \bibinfo{pages}{839982}.
\newblock


\bibitem[Monserrat et~al\mbox{.}(2013)]%
        {monserrat2013notevideo}
\bibfield{author}{\bibinfo{person}{Toni-Jan Keith~Palma Monserrat}, \bibinfo{person}{Shengdong Zhao}, \bibinfo{person}{Kevin McGee}, {and} \bibinfo{person}{Anshul~Vikram Pandey}.} \bibinfo{year}{2013}\natexlab{}.
\newblock \showarticletitle{Notevideo: Facilitating navigation of blackboard-style lecture videos}. In \bibinfo{booktitle}{\emph{Proceedings of the SIGCHI conference on human factors in computing systems}}. \bibinfo{pages}{1139--1148}.
\newblock


\bibitem[Muhsin et~al\mbox{.}(2024)]%
        {muhsin2024review}
\bibfield{author}{\bibinfo{person}{Zahra~J Muhsin}, \bibinfo{person}{Rami Qahwaji}, \bibinfo{person}{Faruque Ghanchi}, {and} \bibinfo{person}{Majid Al-Taee}.} \bibinfo{year}{2024}\natexlab{}.
\newblock \showarticletitle{Review of substitutive assistive tools and technologies for people with visual impairments: recent advancements and prospects}.
\newblock \bibinfo{journal}{\emph{Journal on Multimodal User Interfaces}} \bibinfo{volume}{18}, \bibinfo{number}{1} (\bibinfo{year}{2024}), \bibinfo{pages}{135--156}.
\newblock


\bibitem[Natalie et~al\mbox{.}(2024)]%
        {natalie2024audio}
\bibfield{author}{\bibinfo{person}{Rosiana Natalie}, \bibinfo{person}{Ruei-Che Chang}, \bibinfo{person}{Smitha Sheshadri}, \bibinfo{person}{Anhong Guo}, {and} \bibinfo{person}{Kotaro Hara}.} \bibinfo{year}{2024}\natexlab{}.
\newblock \showarticletitle{Audio description customization}. In \bibinfo{booktitle}{\emph{Proceedings of the 26th International ACM SIGACCESS Conference on Computers and Accessibility}}. \bibinfo{pages}{1--19}.
\newblock


\bibitem[Natalie et~al\mbox{.}(2021)]%
        {natalie2021efficacy}
\bibfield{author}{\bibinfo{person}{Rosiana Natalie}, \bibinfo{person}{Jolene Loh}, \bibinfo{person}{Huei~Suen Tan}, \bibinfo{person}{Joshua Tseng}, \bibinfo{person}{Ian Luke Yi-Ren Chan}, \bibinfo{person}{Ebrima~H Jarjue}, \bibinfo{person}{Hernisa Kacorri}, {and} \bibinfo{person}{Kotaro Hara}.} \bibinfo{year}{2021}\natexlab{}.
\newblock \showarticletitle{The efficacy of collaborative authoring of video scene descriptions}. In \bibinfo{booktitle}{\emph{Proceedings of the 23rd International ACM SIGACCESS Conference on Computers and Accessibility}}. \bibinfo{pages}{1--15}.
\newblock


\bibitem[Navarrete et~al\mbox{.}(2023)]%
        {navarrete2023videoLearning}
\bibfield{author}{\bibinfo{person}{Evelyn Navarrete}, \bibinfo{person}{Andreas Nehring}, \bibinfo{person}{Sascha Schanze}, \bibinfo{person}{Ralph Ewerth}, {and} \bibinfo{person}{Anett Hoppe}.} \bibinfo{year}{2023}\natexlab{}.
\newblock \showarticletitle{A Closer Look into Recent Video-based Learning Research: {A} Comprehensive Review of Video Characteristics, Tools, Technologies, and Learning Effectiveness}.
\newblock \bibinfo{journal}{\emph{CoRR}}  \bibinfo{volume}{abs/2301.13617} (\bibinfo{year}{2023}).
\newblock
\href{https://doi.org/10.48550/ARXIV.2301.13617}{doi:\nolinkurl{10.48550/ARXIV.2301.13617}}
\showeprint[arXiv]{2301.13617}


\bibitem[Ning et~al\mbox{.}(2024)]%
        {ning2024spica}
\bibfield{author}{\bibinfo{person}{Zheng Ning}, \bibinfo{person}{Brianna~L Wimer}, \bibinfo{person}{Kaiwen Jiang}, \bibinfo{person}{Keyi Chen}, \bibinfo{person}{Jerrick Ban}, \bibinfo{person}{Yapeng Tian}, \bibinfo{person}{Yuhang Zhao}, {and} \bibinfo{person}{Toby Jia-Jun Li}.} \bibinfo{year}{2024}\natexlab{}.
\newblock \showarticletitle{SPICA: Interactive Video Content Exploration through Augmented Audio Descriptions for Blind or Low-Vision Viewers}. In \bibinfo{booktitle}{\emph{Proceedings of the CHI Conference on Human Factors in Computing Systems}}. \bibinfo{pages}{1--18}.
\newblock


\bibitem[of~the Blind(nd)]%
        {acb_audio_description_guidelines}
\bibfield{author}{\bibinfo{person}{American~Council of~the Blind}.} \bibinfo{year}{n.d.}\natexlab{}.
\newblock \bibinfo{title}{Audio Description Project, Guidelines for Audio Describers}.
\newblock \bibinfo{howpublished}{\url{https://www.acb.org/adp/guidelines.html}}.
\newblock
\newblock
\shownote{Accessed: 2024-08-10}.


\bibitem[Packer et~al\mbox{.}(2015)]%
        {packer2015overview}
\bibfield{author}{\bibinfo{person}{Jaclyn Packer}, \bibinfo{person}{Katie Vizenor}, {and} \bibinfo{person}{Joshua~A Miele}.} \bibinfo{year}{2015}\natexlab{}.
\newblock \showarticletitle{An overview of video description: history, benefits, and guidelines}.
\newblock \bibinfo{journal}{\emph{Journal of Visual Impairment \& Blindness}} \bibinfo{volume}{109}, \bibinfo{number}{2} (\bibinfo{year}{2015}), \bibinfo{pages}{83--93}.
\newblock


\bibitem[Parveen and Shah(2021)]%
        {parveen2021motion}
\bibfield{author}{\bibinfo{person}{Suraiya Parveen} {and} \bibinfo{person}{Javeria Shah}.} \bibinfo{year}{2021}\natexlab{}.
\newblock \showarticletitle{A motion detection system in python and opencv}. In \bibinfo{booktitle}{\emph{2021 third international conference on intelligent communication technologies and virtual mobile networks (ICICV)}}. IEEE, \bibinfo{publisher}{IEEE}, \bibinfo{address}{Virtual Conference}, \bibinfo{pages}{1378--1382}.
\newblock


\bibitem[Pavel et~al\mbox{.}(2020)]%
        {pavel2020rescribe}
\bibfield{author}{\bibinfo{person}{Amy Pavel}, \bibinfo{person}{Gabriel Reyes}, {and} \bibinfo{person}{Jeffrey~P Bigham}.} \bibinfo{year}{2020}\natexlab{}.
\newblock \showarticletitle{Rescribe: Authoring and automatically editing audio descriptions}. In \bibinfo{booktitle}{\emph{Proceedings of the 33rd Annual ACM Symposium on User Interface Software and Technology}}. \bibinfo{pages}{747--759}.
\newblock


\bibitem[Peng et~al\mbox{.}(2021a)]%
        {peng2021Slidecho}
\bibfield{author}{\bibinfo{person}{Yi{-}Hao Peng}, \bibinfo{person}{Jeffrey~P. Bigham}, {and} \bibinfo{person}{Amy Pavel}.} \bibinfo{year}{2021}\natexlab{a}.
\newblock \showarticletitle{Slidecho: Flexible Non-Visual Exploration of Presentation Videos}. In \bibinfo{booktitle}{\emph{{ASSETS} '21: The 23rd International {ACM} {SIGACCESS} Conference on Computers and Accessibility, Virtual Event, USA, October 18-22, 2021}}, \bibfield{editor}{\bibinfo{person}{Jonathan Lazar}, \bibinfo{person}{Jinjuan~Heidi Feng}, {and} \bibinfo{person}{Faustina Hwang}} (Eds.). \bibinfo{publisher}{{ACM}}, \bibinfo{pages}{24:1--24:12}.
\newblock
\href{https://doi.org/10.1145/3441852.3471234}{doi:\nolinkurl{10.1145/3441852.3471234}}


\bibitem[Peng et~al\mbox{.}(2021b)]%
        {peng2021say}
\bibfield{author}{\bibinfo{person}{Yi-Hao Peng}, \bibinfo{person}{JiWoong Jang}, \bibinfo{person}{Jeffrey~P Bigham}, {and} \bibinfo{person}{Amy Pavel}.} \bibinfo{year}{2021}\natexlab{b}.
\newblock \showarticletitle{Say it all: Feedback for improving non-visual presentation accessibility}. In \bibinfo{booktitle}{\emph{Proceedings of the 2021 CHI Conference on Human Factors in Computing Systems}}. \bibinfo{pages}{1--12}.
\newblock


\bibitem[Prakash et~al\mbox{.}(2024)]%
        {prakash2024understanding}
\bibfield{author}{\bibinfo{person}{Yash Prakash}, \bibinfo{person}{Akshay Kolgar~Nayak}, \bibinfo{person}{Sampath Jayarathna}, \bibinfo{person}{Hae-Na Lee}, {and} \bibinfo{person}{Vikas Ashok}.} \bibinfo{year}{2024}\natexlab{}.
\newblock \showarticletitle{Understanding Low Vision Graphical Perception of Bar Charts}. In \bibinfo{booktitle}{\emph{Proceedings of the 26th International ACM SIGACCESS Conference on Computers and Accessibility}}. \bibinfo{pages}{1--10}.
\newblock


\bibitem[Projects(2024)]%
        {flask}
\bibfield{author}{\bibinfo{person}{Pallets Projects}.} \bibinfo{year}{2024}\natexlab{}.
\newblock \bibinfo{title}{Flask - A micro web framework for Python}.
\newblock \bibinfo{howpublished}{\url{https://flask.palletsprojects.com}}.
\newblock
\newblock
\shownote{Accessed: 2024-09-12}.


\bibitem[Radford et~al\mbox{.}(2023)]%
        {whisper}
\bibfield{author}{\bibinfo{person}{Alec Radford}, \bibinfo{person}{Jong~Wook Kim}, \bibinfo{person}{Tao Xu}, \bibinfo{person}{Greg Brockman}, \bibinfo{person}{Christine McLeavey}, {and} \bibinfo{person}{Ilya Sutskever}.} \bibinfo{year}{2023}\natexlab{}.
\newblock \showarticletitle{Robust speech recognition via large-scale weak supervision}. In \bibinfo{booktitle}{\emph{International conference on machine learning}}. PMLR, \bibinfo{pages}{28492--28518}.
\newblock


\bibitem[Ram et~al\mbox{.}(2023)]%
        {ram2023vidadapter}
\bibfield{author}{\bibinfo{person}{Ashwin Ram}, \bibinfo{person}{Han Xiao}, \bibinfo{person}{Shengdong Zhao}, {and} \bibinfo{person}{Chi-Wing Fu}.} \bibinfo{year}{2023}\natexlab{}.
\newblock \showarticletitle{VidAdapter: Adapting Blackboard-Style Videos for Ubiquitous Viewing}.
\newblock \bibinfo{journal}{\emph{Proc. ACM Interact. Mob. Wearable Ubiquitous Technol.}} \bibinfo{volume}{7}, \bibinfo{number}{3}, Article \bibinfo{articleno}{119} (\bibinfo{date}{sep} \bibinfo{year}{2023}), \bibinfo{numpages}{19}~pages.
\newblock
\href{https://doi.org/10.1145/3610928}{doi:\nolinkurl{10.1145/3610928}}


\bibitem[Sackl et~al\mbox{.}(2020)]%
        {sackl2020ensuring}
\bibfield{author}{\bibinfo{person}{Andreas Sackl}, \bibinfo{person}{Franziska Graf}, \bibinfo{person}{Raimund Schatz}, {and} \bibinfo{person}{Manfred Tscheligi}.} \bibinfo{year}{2020}\natexlab{}.
\newblock \showarticletitle{Ensuring accessibility: Individual video playback enhancements for low vision users}. In \bibinfo{booktitle}{\emph{Proceedings of the 22nd International ACM SIGACCESS Conference on Computers and Accessibility}}. \bibinfo{pages}{1--4}.
\newblock


\bibitem[Schaadhardt et~al\mbox{.}(2021)]%
        {schaadhardt2021understanding}
\bibfield{author}{\bibinfo{person}{Anastasia Schaadhardt}, \bibinfo{person}{Alexis Hiniker}, {and} \bibinfo{person}{Jacob~O Wobbrock}.} \bibinfo{year}{2021}\natexlab{}.
\newblock \showarticletitle{Understanding blind screen-reader users’ experiences of digital artboards}. In \bibinfo{booktitle}{\emph{Proceedings of the 2021 CHI Conference on Human Factors in Computing Systems}}. \bibinfo{pages}{1--19}.
\newblock


\bibitem[Sechayk et~al\mbox{.}(2024)]%
        {sechayk2024smartLearn}
\bibfield{author}{\bibinfo{person}{Yotam Sechayk}, \bibinfo{person}{Ariel Shamir}, {and} \bibinfo{person}{Takeo Igarashi}.} \bibinfo{year}{2024}\natexlab{}.
\newblock \showarticletitle{SmartLearn: Visual-Temporal Accessibility for Slide-based e-learning Videos}. In \bibinfo{booktitle}{\emph{Extended Abstracts of the {CHI} Conference on Human Factors in Computing Systems, {CHI} {EA} 2024, Honolulu, HI, USA, May 11-16, 2024}}, \bibfield{editor}{\bibinfo{person}{Florian~'Floyd' Mueller}, \bibinfo{person}{Penny Kyburz}, \bibinfo{person}{Julie~R. Williamson}, {and} \bibinfo{person}{Corina Sas}} (Eds.). \bibinfo{publisher}{{ACM}}, \bibinfo{pages}{294:1--294:11}.
\newblock
\href{https://doi.org/10.1145/3613905.3650883}{doi:\nolinkurl{10.1145/3613905.3650883}}


\bibitem[Snyder(2005)]%
        {snyder2005audio}
\bibfield{author}{\bibinfo{person}{Joel Snyder}.} \bibinfo{year}{2005}\natexlab{}.
\newblock \showarticletitle{Audio description: The visual made verbal}. In \bibinfo{booktitle}{\emph{International congress series}}, Vol.~\bibinfo{volume}{1282}. Elsevier, \bibinfo{pages}{935--939}.
\newblock


\bibitem[Stangl et~al\mbox{.}(2023)]%
        {stangl2023potential}
\bibfield{author}{\bibinfo{person}{Abigale Stangl}, \bibinfo{person}{Shasta Ihorn}, \bibinfo{person}{Yue-Ting Siu}, \bibinfo{person}{Aditya Bodi}, \bibinfo{person}{Mar Castanon}, \bibinfo{person}{Lothar~D Narins}, {and} \bibinfo{person}{Ilmi Yoon}.} \bibinfo{year}{2023}\natexlab{}.
\newblock \showarticletitle{The Potential of a Visual Dialogue Agent In a Tandem Automated Audio Description System for Videos}. In \bibinfo{booktitle}{\emph{Proceedings of the 25th International ACM SIGACCESS Conference on Computers and Accessibility}}. \bibinfo{pages}{1--17}.
\newblock


\bibitem[Stangl et~al\mbox{.}(2021)]%
        {stangl2021going}
\bibfield{author}{\bibinfo{person}{Abigale Stangl}, \bibinfo{person}{Nitin Verma}, \bibinfo{person}{Kenneth~R. Fleischmann}, \bibinfo{person}{Meredith~Ringel Morris}, {and} \bibinfo{person}{Danna Gurari}.} \bibinfo{year}{2021}\natexlab{}.
\newblock \showarticletitle{Going Beyond One-Size-Fits-All Image Descriptions to Satisfy the Information Wants of People Who are Blind or Have Low Vision}. In \bibinfo{booktitle}{\emph{Proceedings of the 23rd International ACM SIGACCESS Conference on Computers and Accessibility}} (Virtual Event, USA) \emph{(\bibinfo{series}{ASSETS '21})}. \bibinfo{publisher}{Association for Computing Machinery}, \bibinfo{address}{New York, NY, USA}, Article \bibinfo{articleno}{16}, \bibinfo{numpages}{15}~pages.
\newblock
\showISBNx{9781450383066}
\href{https://doi.org/10.1145/3441852.3471233}{doi:\nolinkurl{10.1145/3441852.3471233}}


\bibitem[Stearns et~al\mbox{.}(2018)]%
        {stearns2018design}
\bibfield{author}{\bibinfo{person}{Lee Stearns}, \bibinfo{person}{Leah Findlater}, {and} \bibinfo{person}{Jon~E Froehlich}.} \bibinfo{year}{2018}\natexlab{}.
\newblock \showarticletitle{Design of an augmented reality magnification aid for low vision users}. In \bibinfo{booktitle}{\emph{Proceedings of the 20th international ACM SIGACCESS conference on computers and accessibility}}. \bibinfo{pages}{28--39}.
\newblock


\bibitem[Suzuki et~al\mbox{.}(1985)]%
        {suzuki1985topological}
\bibfield{author}{\bibinfo{person}{Satoshi Suzuki} {et~al\mbox{.}}} \bibinfo{year}{1985}\natexlab{}.
\newblock \showarticletitle{Topological structural analysis of digitized binary images by border following}.
\newblock \bibinfo{journal}{\emph{Computer vision, graphics, and image processing}} \bibinfo{volume}{30}, \bibinfo{number}{1} (\bibinfo{year}{1985}), \bibinfo{pages}{32--46}.
\newblock


\bibitem[Szpiro et~al\mbox{.}(2016)]%
        {szpiro2016people}
\bibfield{author}{\bibinfo{person}{Sarit Felicia~Anais Szpiro}, \bibinfo{person}{Shafeka Hashash}, \bibinfo{person}{Yuhang Zhao}, {and} \bibinfo{person}{Shiri Azenkot}.} \bibinfo{year}{2016}\natexlab{}.
\newblock \showarticletitle{How people with low vision access computing devices: Understanding challenges and opportunities}. In \bibinfo{booktitle}{\emph{Proceedings of the 18th International ACM SIGACCESS Conference on Computers and Accessibility}}. \bibinfo{pages}{171--180}.
\newblock


\bibitem[Tang et~al\mbox{.}(2023)]%
        {tang2023screen}
\bibfield{author}{\bibinfo{person}{Meini Tang}, \bibinfo{person}{Roberto Manduchi}, \bibinfo{person}{Susana Chung}, {and} \bibinfo{person}{Raquel Prado}.} \bibinfo{year}{2023}\natexlab{}.
\newblock \showarticletitle{Screen Magnification for Readers with Low Vision: A Study on Usability and Performance}. In \bibinfo{booktitle}{\emph{Proceedings of the 25th International ACM SIGACCESS Conference on Computers and Accessibility}}. \bibinfo{pages}{1--15}.
\newblock


\bibitem[ThatsEngineering(2024)]%
        {v3}
\bibfield{author}{\bibinfo{person}{ThatsEngineering}.} \bibinfo{year}{2024}\natexlab{}.
\newblock \bibinfo{title}{Frame Assignment For Robotic Manipulators - Direct Kinematics I}.
\newblock \bibinfo{howpublished}{\url{https://www.youtube.com/watch?v=fNIyNF87q9I}}.
\newblock
\newblock
\shownote{Accessed: 2024-09-12}.


\bibitem[Thompson(2024)]%
        {v4}
\bibfield{author}{\bibinfo{person}{Dr.~Chris Thompson}.} \bibinfo{year}{2024}\natexlab{}.
\newblock \bibinfo{title}{Introduction to Neuroscience 2: Lecture 26: Neuroethology}.
\newblock \bibinfo{howpublished}{\url{https://www.youtube.com/watch?v=7pQyw1rHEEg}}.
\newblock
\newblock
\shownote{Accessed: 2024-09-12}.


\bibitem[Tsutsui et~al\mbox{.}(2024)]%
        {TsutsuiYZSTO24}
\bibfield{author}{\bibinfo{person}{Ayaka Tsutsui}, \bibinfo{person}{Kenta Yamamoto}, \bibinfo{person}{Yinan Zhao}, \bibinfo{person}{Ippei Suzuki}, \bibinfo{person}{Kengo Tanaka}, {and} \bibinfo{person}{Yoichi Ochiai}.} \bibinfo{year}{2024}\natexlab{}.
\newblock \showarticletitle{Low Vision Boxing: Participatory Design of Adaptive Kickboxing Experiences with Low Vision Person}. In \bibinfo{booktitle}{\emph{The 26th International {ACM} {SIGACCESS} Conference on Computers and Accessibility, {ASSETS} 2024, St. John's, NL, Canada, October 27-30, 2024}}, \bibfield{editor}{\bibinfo{person}{David~R. Flatla}, \bibinfo{person}{Faustina Hwang}, \bibinfo{person}{Tiago Jo{\~{a}}o~Vieira Guerreiro}, {and} \bibinfo{person}{Robin Brewer}} (Eds.). \bibinfo{publisher}{{ACM}}, \bibinfo{pages}{15:1--15:18}.
\newblock
\href{https://doi.org/10.1145/3663548.3675619}{doi:\nolinkurl{10.1145/3663548.3675619}}


\bibitem[Tutor(2024)]%
        {v1}
\bibfield{author}{\bibinfo{person}{The Organic~Chemistry Tutor}.} \bibinfo{year}{2024}\natexlab{}.
\newblock \bibinfo{title}{Biology - Intro to Cell Structure - Quick Review!}
\newblock \bibinfo{howpublished}{\url{https://www.youtube.com/watch?v=vwAJ8ByQH2U}}.
\newblock
\newblock
\shownote{Accessed: 2024-09-12}.


\bibitem[Wang et~al\mbox{.}(2024a)]%
        {WangPHKZM024}
\bibfield{author}{\bibinfo{person}{Ru Wang}, \bibinfo{person}{Zach Potter}, \bibinfo{person}{Yun Ho}, \bibinfo{person}{Daniel Killough}, \bibinfo{person}{Linxiu Zeng}, \bibinfo{person}{Sanbrita Mondal}, {and} \bibinfo{person}{Yuhang Zhao}.} \bibinfo{year}{2024}\natexlab{a}.
\newblock \showarticletitle{GazePrompt: Enhancing Low Vision People's Reading Experience with Gaze-Aware Augmentations}. In \bibinfo{booktitle}{\emph{Proceedings of the {CHI} Conference on Human Factors in Computing Systems, {CHI} 2024, Honolulu, HI, USA, May 11-16, 2024}}, \bibfield{editor}{\bibinfo{person}{Florian~'Floyd' Mueller}, \bibinfo{person}{Penny Kyburz}, \bibinfo{person}{Julie~R. Williamson}, \bibinfo{person}{Corina Sas}, \bibinfo{person}{Max~L. Wilson}, \bibinfo{person}{Phoebe O.~Toups Dugas}, {and} \bibinfo{person}{Irina Shklovski}} (Eds.). \bibinfo{publisher}{{ACM}}, \bibinfo{pages}{894:1--894:17}.
\newblock
\href{https://doi.org/10.1145/3613904.3642878}{doi:\nolinkurl{10.1145/3613904.3642878}}


\bibitem[Wang et~al\mbox{.}(2023)]%
        {wang2023understanding}
\bibfield{author}{\bibinfo{person}{Ru Wang}, \bibinfo{person}{Linxiu Zeng}, \bibinfo{person}{Xinyong Zhang}, \bibinfo{person}{Sanbrita Mondal}, {and} \bibinfo{person}{Yuhang Zhao}.} \bibinfo{year}{2023}\natexlab{}.
\newblock \showarticletitle{Understanding how low vision people read using eye tracking}. In \bibinfo{booktitle}{\emph{Proceedings of the 2023 CHI Conference on Human Factors in Computing Systems}}. \bibinfo{publisher}{ACM}, \bibinfo{address}{New York, NY, USA}, \bibinfo{pages}{1--17}.
\newblock


\bibitem[Wang et~al\mbox{.}(2021)]%
        {wang2021toward}
\bibfield{author}{\bibinfo{person}{Yujia Wang}, \bibinfo{person}{Wei Liang}, \bibinfo{person}{Haikun Huang}, \bibinfo{person}{Yongqi Zhang}, \bibinfo{person}{Dingzeyu Li}, {and} \bibinfo{person}{Lap-Fai Yu}.} \bibinfo{year}{2021}\natexlab{}.
\newblock \showarticletitle{Toward automatic audio description generation for accessible videos}. In \bibinfo{booktitle}{\emph{Proceedings of the 2021 CHI Conference on Human Factors in Computing Systems}}. \bibinfo{pages}{1--12}.
\newblock


\bibitem[Wang et~al\mbox{.}(2024b)]%
        {wang2024low}
\bibfield{author}{\bibinfo{person}{Yanan Wang}, \bibinfo{person}{Yuhang Zhao}, {and} \bibinfo{person}{Yea-Seul Kim}.} \bibinfo{year}{2024}\natexlab{b}.
\newblock \showarticletitle{How Do Low-Vision Individuals Experience Information Visualization?}. In \bibinfo{booktitle}{\emph{Proceedings of the CHI Conference on Human Factors in Computing Systems}}. \bibinfo{pages}{1--15}.
\newblock


\bibitem[Yip et~al\mbox{.}(2021)]%
        {yip2021visionary}
\bibfield{author}{\bibinfo{person}{Carmen Yip}, \bibinfo{person}{Jie~Mi Chong}, \bibinfo{person}{Sin~Yee Kwek}, \bibinfo{person}{Yong Wang}, {and} \bibinfo{person}{Kotaro Hara}.} \bibinfo{year}{2021}\natexlab{}.
\newblock \showarticletitle{Visionary Caption: Improving the Accessibility of Presentation Slides through Highlighting Visualization}. In \bibinfo{booktitle}{\emph{Proceedings of the 23rd International ACM SIGACCESS Conference on Computers and Accessibility}}. \bibinfo{pages}{1--4}.
\newblock


\bibitem[{YouTube}({[n.\,d.]})]%
        {youtube}
\bibfield{author}{\bibinfo{person}{{YouTube}}.} \bibinfo{year}{[n.\,d.]}\natexlab{}.
\newblock \bibinfo{title}{YouTube}.
\newblock \bibinfo{howpublished}{\url{https://www.youtube.com/}}.
\newblock
\newblock
\shownote{Accessed: 2024-08-10}.


\bibitem[Yuksel et~al\mbox{.}(2020)]%
        {yuksel2020human}
\bibfield{author}{\bibinfo{person}{Beste~F Yuksel}, \bibinfo{person}{Pooyan Fazli}, \bibinfo{person}{Umang Mathur}, \bibinfo{person}{Vaishali Bisht}, \bibinfo{person}{Soo~Jung Kim}, \bibinfo{person}{Joshua~Junhee Lee}, \bibinfo{person}{Seung~Jung Jin}, \bibinfo{person}{Yue-Ting Siu}, \bibinfo{person}{Joshua~A Miele}, {and} \bibinfo{person}{Ilmi Yoon}.} \bibinfo{year}{2020}\natexlab{}.
\newblock \showarticletitle{Human-in-the-loop machine learning to increase video accessibility for visually impaired and blind users}. In \bibinfo{booktitle}{\emph{Proceedings of the 2020 ACM Designing Interactive Systems Conference}}. \bibinfo{pages}{47--60}.
\newblock


\bibitem[Zhao et~al\mbox{.}(2019)]%
        {ZhaoKCFA19}
\bibfield{author}{\bibinfo{person}{Yuhang Zhao}, \bibinfo{person}{Elizabeth Kupferstein}, \bibinfo{person}{Brenda~Veronica Castro}, \bibinfo{person}{Steven Feiner}, {and} \bibinfo{person}{Shiri Azenkot}.} \bibinfo{year}{2019}\natexlab{}.
\newblock \showarticletitle{Designing {AR} Visualizations to Facilitate Stair Navigation for People with Low Vision}. In \bibinfo{booktitle}{\emph{Proceedings of the 32nd Annual {ACM} Symposium on User Interface Software and Technology, {UIST} 2019, New Orleans, LA, USA, October 20-23, 2019}}, \bibfield{editor}{\bibinfo{person}{Fran{\c{c}}ois Guimbreti{\`{e}}re}, \bibinfo{person}{Michael~S. Bernstein}, {and} \bibinfo{person}{Katharina Reinecke}} (Eds.). \bibinfo{publisher}{{ACM}}, \bibinfo{pages}{387--402}.
\newblock
\href{https://doi.org/10.1145/3332165.3347906}{doi:\nolinkurl{10.1145/3332165.3347906}}


\bibitem[Zhao et~al\mbox{.}(2020a)]%
        {ZhaoKRFA20}
\bibfield{author}{\bibinfo{person}{Yuhang Zhao}, \bibinfo{person}{Elizabeth Kupferstein}, \bibinfo{person}{Hathaitorn Rojnirun}, \bibinfo{person}{Leah Findlater}, {and} \bibinfo{person}{Shiri Azenkot}.} \bibinfo{year}{2020}\natexlab{a}.
\newblock \showarticletitle{The Effectiveness of Visual and Audio Wayfinding Guidance on Smartglasses for People with Low Vision}. In \bibinfo{booktitle}{\emph{{CHI} '20: {CHI} Conference on Human Factors in Computing Systems, Honolulu, HI, USA, April 25-30, 2020}}, \bibfield{editor}{\bibinfo{person}{Regina Bernhaupt}, \bibinfo{person}{Florian~'Floyd' Mueller}, \bibinfo{person}{David Verweij}, \bibinfo{person}{Josh Andres}, \bibinfo{person}{Joanna McGrenere}, \bibinfo{person}{Andy Cockburn}, \bibinfo{person}{Ignacio Avellino}, \bibinfo{person}{Alix Goguey}, \bibinfo{person}{Pernille Bj{\o}n}, \bibinfo{person}{Shengdong Zhao}, \bibinfo{person}{Briane~Paul Samson}, {and} \bibinfo{person}{Rafal Kocielnik}} (Eds.). \bibinfo{publisher}{{ACM}}, \bibinfo{pages}{1--14}.
\newblock
\href{https://doi.org/10.1145/3313831.3376516}{doi:\nolinkurl{10.1145/3313831.3376516}}


\bibitem[Zhao et~al\mbox{.}(2016)]%
        {zhao2016cuesee}
\bibfield{author}{\bibinfo{person}{Yuhang Zhao}, \bibinfo{person}{Sarit Szpiro}, \bibinfo{person}{Jonathan Knighten}, {and} \bibinfo{person}{Shiri Azenkot}.} \bibinfo{year}{2016}\natexlab{}.
\newblock \showarticletitle{CueSee: exploring visual cues for people with low vision to facilitate a visual search task}. In \bibinfo{booktitle}{\emph{Proceedings of the 2016 ACM International Joint Conference on Pervasive and Ubiquitous Computing}} (Heidelberg, Germany) \emph{(\bibinfo{series}{UbiComp '16})}. \bibinfo{publisher}{Association for Computing Machinery}, \bibinfo{address}{New York, NY, USA}, \bibinfo{pages}{73–84}.
\newblock
\showISBNx{9781450344616}
\href{https://doi.org/10.1145/2971648.2971730}{doi:\nolinkurl{10.1145/2971648.2971730}}


\bibitem[Zhao et~al\mbox{.}(2020b)]%
        {zhao2020Designing}
\bibfield{author}{\bibinfo{person}{Yuhang Zhao}, \bibinfo{person}{Sarit Szpiro}, \bibinfo{person}{Lei Shi}, {and} \bibinfo{person}{Shiri Azenkot}.} \bibinfo{year}{2020}\natexlab{b}.
\newblock \showarticletitle{Designing and Evaluating a Customizable Head-mounted Vision Enhancement System for People with Low Vision}.
\newblock \bibinfo{journal}{\emph{{ACM} Trans. Access. Comput.}} \bibinfo{volume}{12}, \bibinfo{number}{4} (\bibinfo{year}{2020}), \bibinfo{pages}{15:1--15:46}.
\newblock
\href{https://doi.org/10.1145/3361866}{doi:\nolinkurl{10.1145/3361866}}


\bibitem[{Zoom}({[n.\,d.]})]%
        {zoom}
\bibfield{author}{\bibinfo{person}{{Zoom}}.} \bibinfo{year}{[n.\,d.]}\natexlab{}.
\newblock \bibinfo{title}{Zoom}.
\newblock \bibinfo{howpublished}{\url{https://zoom.us/}}.
\newblock
\newblock
\shownote{Accessed: 2024-08-10}.


\end{thebibliography}

\appendix

\section{Empirical Evaluation of Visual Activities in Presentation Videos}
\label{apdx:presentations}

To better understand how visual activities appear in educational presentation videos in the wild, we conducted an analysis of 300 YouTube~\cite{youtube} videos. We examined the presence of visual activities—namely pointing, sketching, and marking—across a range of academic disciplines.

\subsection{Method}
Following a sampling approach inspired by prior work~\cite{peng2021say}, we selected 6 random disciplines from each of five domain categories (applied sciences, formal sciences, natural sciences, social sciences, and humanities), based on Wikipedia's Outline of Academic Disciplines.\footnote{\url{https://en.wikipedia.org/wiki/Outline_of_academic_disciplines}} This yielded a total of 30 disciplines.

We recruited 6 participants to perform the analysis. Each was assigned a randomized subset of disciplines. For each discipline, participants performed YouTube searches using two queries: \verb|lesson| and \verb|presentation|. From each query, the first 5 unique videos over 4 minutes in length that included screen-shared slides were selected, resulting in 10 videos per discipline. If no suitable videos are found, the search was broadened to a higher-level category.
Participants rated each video on a 5-point Likert scale for the frequency of three visual activity types: pointing, sketching, and marking.

\subsection{Results}

\begin{figure}[htbp]
    \centering
    \includegraphics[width=0.75\linewidth]{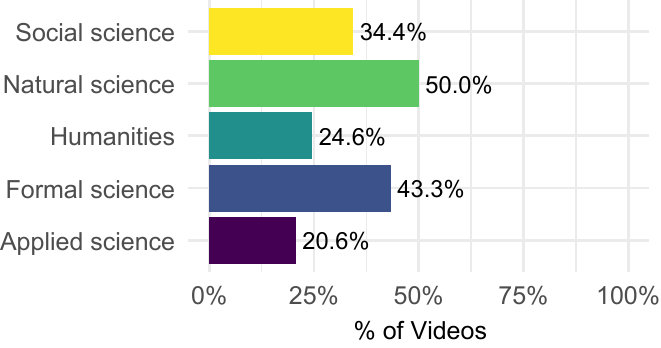}
    \caption{The ratio of visual activities per domain. Any presentation video with any of: pointing, sketching, or marking activities is considered to have visual activities.}
    \Description{Bar chart showing the ratio of any visual activities per domain with usage percentages on the y-axis and various domains on the x-axis. The percentages are as follows: Applied Science (20.6\%), Formal Science (43.3\%), Humanities (24.6\%), Natural Science (50.0\%), and Social Science (34.4\%).}
    \label{fig:empirical_domain_ratio}
\end{figure}

\subsubsection{Overall Usage of Visual Activities}
Visual activities were observed across all domains, though frequency varied. For example, videos in Natural Sciences and Formal Sciences featured visual activities in 50\% and 43.3\% of cases, respectively. See Figure~\ref{fig:empirical_domain_ratio} for full domain-level breakdowns.

\begin{figure}[htbp]
    \centering
    \includegraphics[width=0.95\linewidth]{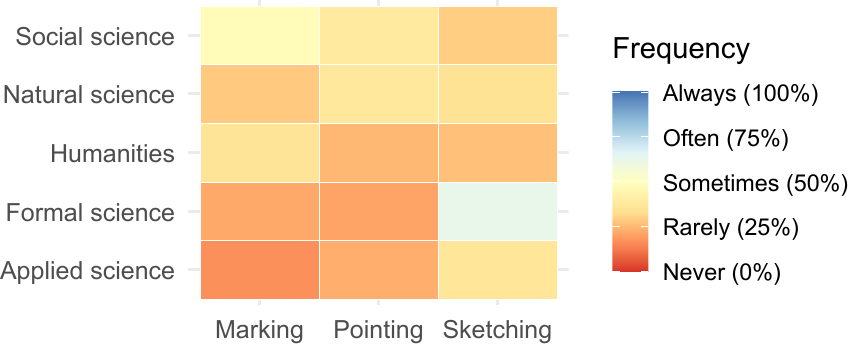}
    \caption{The frequency of visual activities within videos per domain, excluding videos without visual activities. Participants reviewed each video, noting how often instructors use pointing, marking or sketching. The reviewing process included skimming the video using the timeline.}
    \Description{Heatmap displaying the frequency of Activities (when used) across different domains (Applied Science, Formal Science, Humanities, Natural Science, Social Science) and activity types (Sketching, Pointing, Marking). The color gradient represents frequency: Always (dark blue) to Never (red). Most activities in various domains are depicted in yellow-orange hues, indicating they are used Rarely or Sometimes, with one instances of "Often" appearing on "sketching" for "formal sciences."}
    \label{fig:empirical_activity_frequency}
\end{figure}

\subsubsection{Frequency of Activities within Videos by Domain}
We examined the frequency within videos of each activity type for each domain, excluding videos that do not have any activity present.  In \textit{Natural Sciences} (50\% of videos) both sketching is used as frequently as pointing, while in \textit{Formal Sciences} (43.3\% of videos), sketching is used more frequently compared to marking and pointing which are less common. In \textit{Social Sciences} (34.4\% of videos), however, marking is used more frequently compared to pointing and sketching.  See Figure~\ref{fig:empirical_activity_frequency} for a detailed breakdown.

\section{Personalization Usage by Low Vision Participants}
\label{apdx:personalization_usage}
\begin{figure}[htbp]
    \centering
    \includegraphics[width=0.7\linewidth]{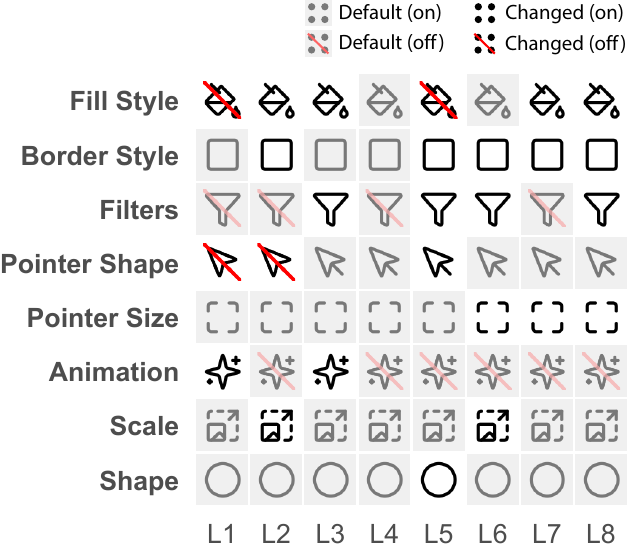}
    \caption{Detailed usage for each LV participant on their modifications of the settings. For each setting we indicate if they kept it as the default or changed it, and if it was on/off.}
    \Description{Matrix that demonstrates for each low vision participants which setting category they changed (or disabled) compared to the default settings.}
    \label{fig:personalization_details}
\end{figure}

\begin{figure}[htbp]
    \centering
    \includegraphics[width=0.7\linewidth]{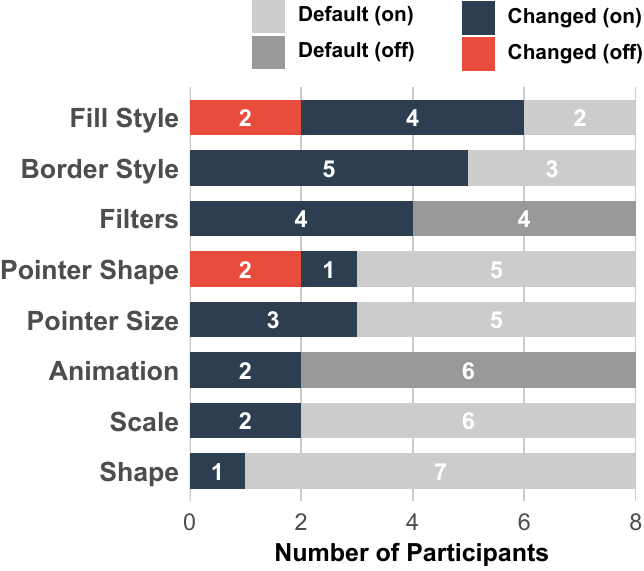}
    \caption{Aggregate of usage per individual setting for LV participants. For each setting we indicate how many users kept it as the default or changed it, and if it was on/off.}
    \Description{Horizontal stacked bar chart. The most changed setting is "fill style" with 6 participants (2 of whom disabled it), then "border style" with 5 participants, "filters" with 4 participants, and more. "Shape" was changed by only one LV participant.}
    \label{fig:personalization_summary}
\end{figure}

In our user study, we provided LV participants with the ability to personalize their experience with VeasyGuide. This included adjusting settings such as fill or border styles, animation styles, and others. See Figure~\ref{fig:personalization_details} for detailed usage by participants, and Figure~\ref{fig:personalization_summary} for an aggregate by setting.

\newpage

\section{Normality Checks for Success Rates and Detection Speeds of Low Vision Participants}
\label{apdx:normality_checks}

\begin{figure}[htbp]
    \centering
    \includegraphics[width=0.8\linewidth]{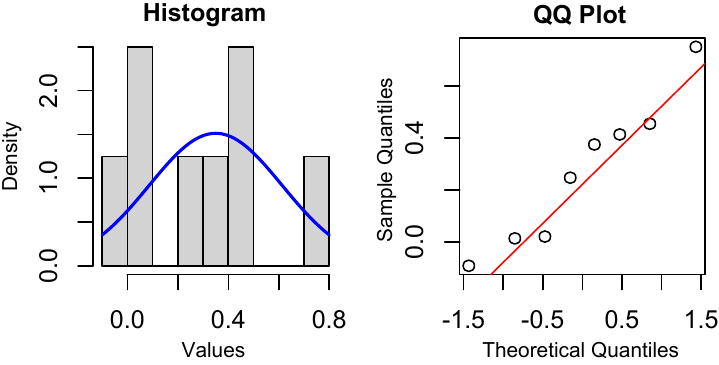}
    \caption{Normality of success rate distribution for LV participants in localization task. Normally distributed.}
    \Description{Histogram showing the distribution of success rates for low vision participants in both conditions, and Q-Q plot indicating how closely the data follows a normal distribution.}
    \label{fig:normality_success_lv}
\end{figure}

\begin{figure}[htbp]
    \centering
    \includegraphics[width=0.8\linewidth]{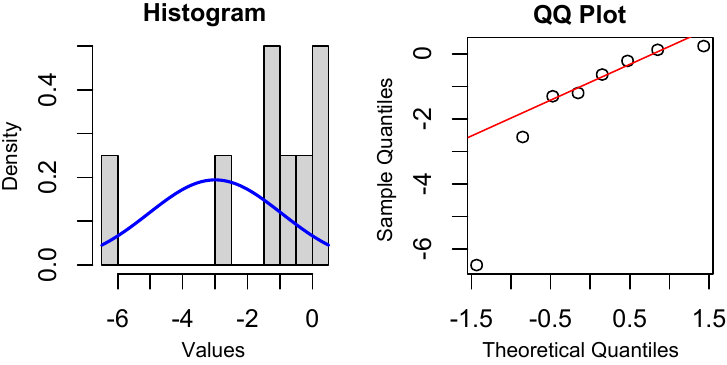}
    \caption{Normality of detection speed distribution for LV participants in localization task. Not normally distributed.}
    \Description{Histogram showing the distribution of detection times for low vision participants in both conditions, and Q-Q plot indicating how closely the data follows a normal distribution.}
    \label{fig:normality_speed_lv}
\end{figure}

As part of our normality assessment, we visualized the distributions using histograms and Q-Q plots (see Figure~\ref{fig:normality_success_lv}--\ref{fig:normality_speed_lv}).

\section{Viewing Task Quiz Completion Time}
\label{apdx:quiz_completion_time}
\begin{figure}[htbp]
    \centering
    \includegraphics[width=0.6\linewidth]{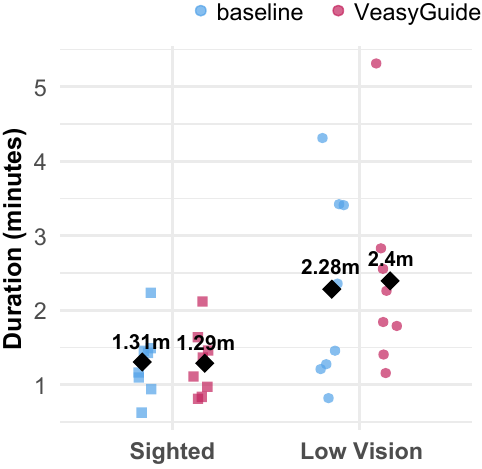}
    \caption{Quiz completion times for both groups are shown, with mean times indicated. No significant differences found.}
    \Description{Scatter plot showing the distribution of quiz completion times for sighted and low vision participants.}
    \label{fig:quiz_completion_time}
\end{figure}

To encourage engagement, participants completed a six-question multiple-choice quiz after each video to assess their understanding. Figure~\ref{fig:quiz_completion_time} shows the distribution of quiz completion times.

\newpage

\section{Success Rates and Detection Speeds Per Video for Low Vision Participants}
\label{apdx:video_user_results}

\begin{figure}[htbp]
    \centering
    \includegraphics[width=0.65\linewidth]{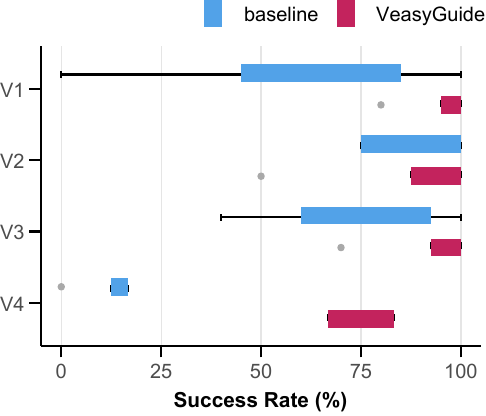}
    \caption{Success rates of LV participants in the localization task per video for both baseline and VeasyGuide conditions.}
    \Description{Bar chart showing the success rates of low vision participants in the localization task per video for both baseline and VeasyGuide conditions. Each video has two bars indicating the success rate in each condition. For VeasyGuide all success rates are higher and with much smaller variance.}
    \label{fig:video_success_lv}
\end{figure}

\begin{figure}[htbp]
    \centering
    \includegraphics[width=0.65\linewidth]{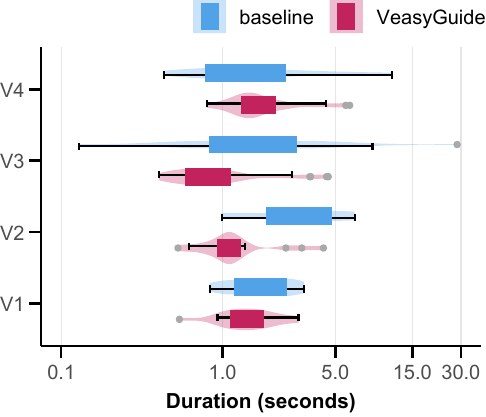}
    \caption{Detection speeds of LV participants in the localization task per video for both baseline and VeasyGuide conditions.}
    \Description{Box plot showing the detection times of low vision participants in the localization task per video for both baseline and VeasyGuide conditions. Each video has two boxes indicating the detection speed in each condition. The mean detection times are similar, however VeasyGuide shows smaller variance.}
    \label{fig:video_speed_lv}
\end{figure}

Results on success rates (Figure~\ref{fig:video_success_lv}) and detection speeds (Figure~\ref{fig:video_speed_lv}) of LV participants aggregated per video in the localization task. We note that V4~\cite{v4} perceived as more challenging for visual search, with \(\ <25\% \) success in the baseline condition.

\clearpage
\onecolumn
\section{Full Localization Task Results}
\label{apdx:full_user_results}
We present the full results of the localization task, including detailed success rates and detection speeds for both LV and sighted participants across all videos (V1--V4).

\begin{table}[h]
    \centering
    \caption{Detailed breakdown of results for success rates and detection speeds from the localization task for both LV and sighted participants in the baseline and VeasyGuide conditions. ``Total,'' is the total number of activities for that condition.}
    \begin{tabular}{lcccccccccccc}
    \toprule
        & \multicolumn{6}{c}{\textbf{baseline}} & \multicolumn{6}{c}{\textbf{VeasyGuide}}                                   \\
    \cmidrule(lr){2-7} \cmidrule(lr){8-13}
        & \multicolumn{2}{c}{Success Rate}      & \multicolumn{4}{c}{Duration (seconds)}
        & \multicolumn{2}{c}{Success Rate}      & \multicolumn{4}{c}{Duration (seconds)}                                    \\
    \cmidrule(lr){2-3} \cmidrule(lr){4-7} \cmidrule(lr){8-9} \cmidrule(lr){10-13}
    PID & Success                               & (total \#)                              & Mean  & SD     & Median & IQR
        & Success                               & (total \#)                              & Mean  & SD     & Median & IQR   \\
    \midrule
    L1  & 0.545                                 & 11                                      & 2.284 & 1.113  & 2.584  & 1.678
        & 1.000                                 & 18                                      & 1.084 & 0.478  & 0.856  & 0.695 \\
    L2  & 0.625                                 & 16                                      & 4.001 & 2.363  & 3.829  & 2.569
        & 1.000                                 & 13                                      & 1.449 & 0.693  & 1.374  & 0.288 \\
    L3  & 0.706                                 & 17                                      & 3.761 & 2.908  & 3.248  & 2.114
        & 0.727                                 & 11                                      & 2.461 & 0.912  & 2.352  & 1.108 \\
    L4  & 0.462                                 & 13                                      & 1.248 & 0.554  & 1.259  & 0.873
        & 0.875                                 & 16                                      & 0.616 & 0.168  & 0.568  & 0.224 \\
    L5  & 0.364                                 & 11                                      & 3.394 & 2.137  & 2.431  & 1.242
        & 0.611                                 & 18                                      & 3.521 & 1.804  & 3.511  & 1.914 \\
    L6  & 0.250                                 & 16                                      & 8.237 & 13.482 & 1.820  & 8.066
        & 1.000                                 & 13                                      & 1.749 & 0.568  & 1.700  & 0.681 \\
    L7  & 1.000                                 & 18                                      & 0.878 & 0.624  & 0.657  & 0.523
        & 0.909                                 & 11                                      & 1.120 & 0.253  & 1.110  & 0.177 \\
    L8  & 0.923                                 & 13                                      & 1.343 & 0.586  & 1.170  & 0.271
        & 0.938                                 & 16                                      & 1.130 & 0.354  & 1.030  & 0.211 \\
    \midrule
    S1  & 1.000                                 & 18                                      & 0.772 & 0.338  & 0.680  & 0.203
        & 1.000                                 & 11                                      & 0.861 & 0.424  & 0.798  & 0.359 \\
    S2  & 1.000                                 & 13                                      & 1.281 & 0.321  & 1.124  & 0.542
        & 1.000                                 & 16                                      & 1.148 & 0.204  & 1.096  & 0.212 \\
    S3  & 1.000                                 & 11                                      & 1.475 & 0.261  & 1.416  & 0.419
        & 1.000                                 & 18                                      & 1.185 & 0.357  & 1.058  & 0.454 \\
    S4  & 1.000                                 & 16                                      & 1.865 & 1.591  & 1.208  & 1.190
        & 1.000                                 & 13                                      & 0.953 & 0.349  & 0.885  & 0.418 \\
    S5  & 0.944                                 & 18                                      & 1.025 & 0.923  & 0.693  & 0.203
        & 1.000                                 & 11                                      & 0.798 & 0.138  & 0.760  & 0.091 \\
    S6  & 1.000                                 & 13                                      & 1.012 & 0.371  & 0.948  & 0.400
        & 1.000                                 & 16                                      & 0.996 & 0.262  & 0.871  & 0.215 \\
    S7  & 1.000                                 & 11                                      & 0.688 & 0.253  & 0.683  & 0.240
        & 1.000                                 & 18                                      & 0.972 & 0.299  & 0.923  & 0.258 \\
    S8  & 1.000                                 & 16                                      & 1.848 & 0.426  & 1.860  & 0.567
        & 1.000                                 & 13                                      & 1.144 & 0.431  & 0.957  & 0.171 \\
    \bottomrule
\end{tabular}
    \label{tab:all_localization_results}
\end{table}


\end{document}